\documentclass[11pt,a4paper]{article}
\usepackage{jheppub}

\usepackage{bbold}

\usepackage[english]{babel}

\newmuskip\pFqmuskip

\newcommand*\pFq[6][8]{%
  \begingroup 
  \pFqmuskip=#1mu\relax
  \mathchardef\normalcomma=\mathcode`,
  \mathcode`\,=\string"8000
  \begingroup\lccode`\~=`\,
  \lowercase{\endgroup\let~}\pFqcomma
  {}_{#2}F_{#3}{\left(\genfrac..{0pt}{}{#4}{#5}\Big | #6\right)}%
  \endgroup
}
\newcommand{\pFqcomma}{{\normalcomma}\mskip\pFqmuskip}

\newcommand{\pmat}{\begin{pmatrix}}
\newcommand{\fpmat}{\end{pmatrix}}
\newcommand{\eq}{\begin{equation}}
\newcommand{\feq}{\end{equation}}
\newcommand{\cas}{\begin{cases}}
\newcommand{\fcas}{\end{cases}}

\newcommand{\eqarray}{\begin{eqnarray}}
\newcommand{\feqarray}{\end{eqnarray}}

\newcommand{\be}{\beta}

\def\be{\begin{equation}}
\def\ee{\end{equation}}
\def\bea{\begin{eqnarray}}
\def\eea{\end{eqnarray}}

\title{\centering{The chaotic emergence of thermalization in highly excited string decays}}

\author{Maurizio Firrotta}
\affiliation[a]{Dipartimento di Fisica, Università  di Roma Tor Vergata, Via della Ricerca Scientifica 1, 00133, Roma, Italy}
\affiliation[b]{INFN sezione di Roma Tor Vergata, 
Via della Ricerca Scientifica 1, 00133 Roma, Italy}

\emailAdd{maurizio.firrotta@gmail.com}

\abstract{We analyse the most general process of a generic highly excited string that decays into a less excited, yet generic, highly excited string emitting a tachyon. We provide a simple and compact analytic description of the decay process which discriminates  between and within the structure of every single microstate of the initial and final highly excited string. Taking into account the random nature of the decay process we extract the energy spectrum of highly excited strings, microstate by microstate, finding a behavior which corresponds to the greybody emission spectrum. In addition, by exploiting the analytic control of the decay process, we identify the origin of thermal effects which are triggered by the chaotic nature of the highly excited string interactions modeled by the microstates structure. 
 }

\makeatletter
\gdef\@fpheader{}
\makeatother

\begin{document}
\maketitle

\section*{Introduction}
The present paper is focused on enlightening the connection between chaos and thermal effects within the physical systems provided by highly excited string (HES) interactions. Motivated by the intriguing interplay between chaos, thermal effects and quantum information \cite{Gibbons:1977mu}-\cite{Shenker:2013pqa}, which are three milestones of black hole (BH) physics, we first used HES as promising candidates of BH states \cite{Susskind:1993ws}-\cite{Sundborg:2000wp} and then we computed their energy spectra. The main goal was to detect a manifest connection between the chaotic behavior of HES interactions and the thermalization of their energy spectra which emerges naturally. In line with past studies on string decays and the produced Hawking radiation \cite{Amati}-\cite{Cornalba:2006hc}, we used and improved the most general process of an HES that decays into an HES emitting a tachyon \cite{Bianchi:2019ywd}\cite{Firrotta:2022cku}, providing an analytic description of the decay process which discriminates between and within the structure of every single microstate of the initial and final HES. Considering the random nature of the decay process we extracted the spectrum of the HES, microstate by microstate, finding a behavior which corresponds to the greybody emission spectrum. In addition, exploiting the analytic control of the decay process, we identified  the origin of thermal effects, finding that they are triggered by the chaotic nature of HES interactions.

The setup we adopted relies on the recent improvement of the Di Vecchia, Del Giudice and Fubini (DDF) formalism \cite{DelGiudice:1971yjh}\cite{Brower:1972wj}, where its spectrum generating algebra was recasted in a manifestly covariant form \cite{Skliros:2011si}\cite{Aldi:2019osr}, for both bosonic string and superstring theories. The possibility of identifying each state of the string spectrum with the associated physical vertex operator gave rise to a wide range of applications: the realization of the scattering of string coherent vertex operators\footnote{A very powerful application of the scattering amplitudes of coherent string states is their nature of generating amplitudes of any desired string states, obtained by a simple derivative projection over mass eigenstates.} \cite{Bianchi:2019ywd}, the non perturbative string footprint in the gravitational wave (GW) signal produced in the merging phase of BHs collision \cite{Addazi:2020obs}, the non perturbative spinning corrections to the electromagnetic wave produced in the collision of heavy sized objects \cite{Aldi:2021zhh}, such as BHs and neutron stars NSs, the two body decay of HES \cite{Firrotta:2022cku} and finally the indications about the chaotic behavior of HES interactions \cite{Gross:2021gsj}\cite{Rosenhaus:2021xhm}. 

About the chaotic behavior of HES interactions, quite recently a novel measure of chaos for scattering amplitudes was proposed \cite{ChaosScatt}, where the behavior of HES amplitudes was compared with the chaotic distribution of the zeros of the Riemann zeta function and the quantum mechanical scattering on a leaky torus, finding a common pattern among them.

The scope of the present paper was to continue the study of the physical applications connected to the possibility of exploring the interactions of the whole tower of string excitations, or string microstates, proceeding beyond the physics of light string states and the first Regge trajectory.  

The paper is organized as follows: in section \ref{se1} we explained the connection between the shape of classical string configurations and the structure of quantum string configurations, in particular we studied the classical string profiles as a function of the number of harmonics and the respective coefficients and we compared their shape with the degenerate quantum string partitions of generic mass levels. We followed the logic that the structure of a quantum string state can be probed through its interaction, we selected the simplest one $i.e$ the decay of HES into two tachyons, in order to preserve the HES structure. After a brief review of the chaotic analysis, developed in \cite{ChaosScatt}, we compared the shape of classical string configurations with the chaotic behavior of the scattering amplitudes relative to the quantum analog of classical string configurations finding a common pattern.

In section \ref{se2} we studied the most general process of a generic HES that decays into a less excited, yet generic, HES emitting a tachyon in the thermalization regime. Exploiting the analytic control of the decay process, we identify the origin of thermal effects with the chaotic structure of the process. Finally we gave a description of how to compute the emission spectrum of HES.

In section \ref{se3} we presented the results of HES spectra, microstate by microstate. In particular we found that a generic string excitation is characterized by a greybody emission, while for the extreme case where only excitations of the first Regge trajectory (FRtj) are considered, the thermal nature is highly suppressed as to be negligible.

In appendix \ref{AApp} we reviewed the computation developed in \cite{Firrotta:2022cku}, and we described the analytical implications of the thermalization regime, or more precisely the regime in which the ratio between the energy of the emitted state and the mass of the decaying state is enough small in such a way that the energy loss of the decaying state is smooth.

\section{Chaos in highly excited string processes}\label{se1}
The aim of this section is to review recent results about the chaotic behavior of HES interactions.

\subsection{Classical string vs quantum string configurations}
\begin{figure}[h!] 
\centering
\includegraphics[scale=0.5]{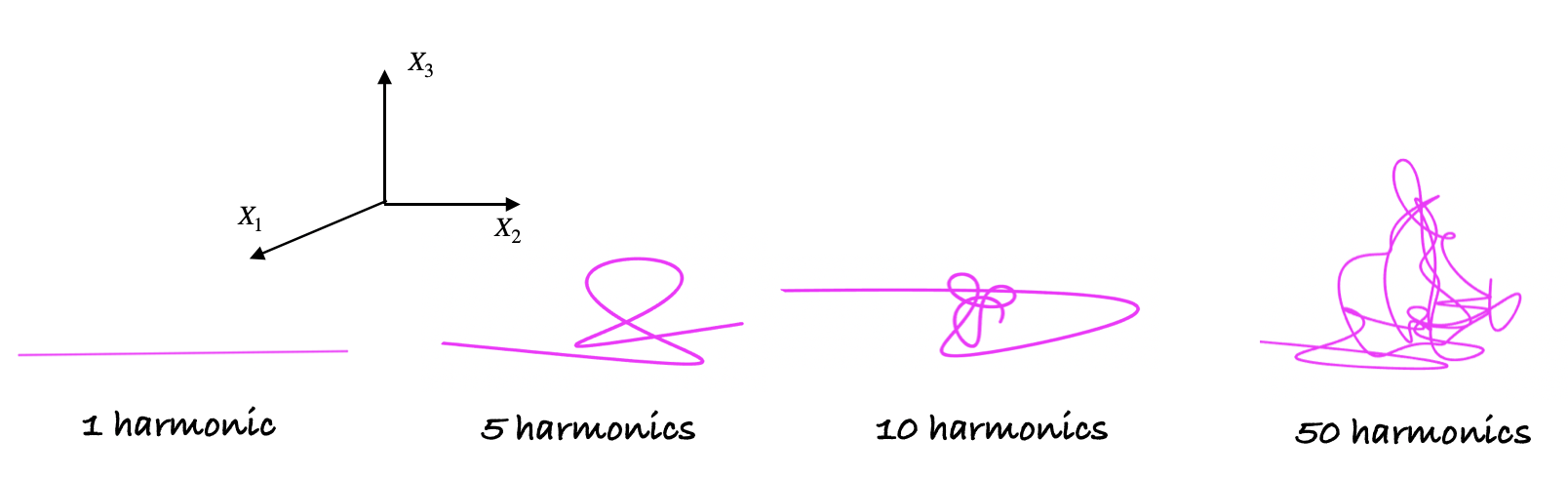}
\caption{Examples of 3D string profiles for different combinations of harmonics  $n^{*}=1,5,10,50$ with uniformly distributed random parameters $a_{n}^{j}\in (0,1)$. } \label{f1}
\end{figure}

Classical three dimensional\footnote{In the specific case of classical string, it was chosen $j=1,2,3$ in order to plot 3D profiles, but in general $j=1,...,D-2$.} bosonic open string profiles with Neumann boundary conditions, in the temporal gauge ($X^{0}=t$), are given by \cite{Green:1987sp}

\be\label{fbk1}
X^{j}_{\{a_{n}\}}(\sigma,\tau)=\sum_{n=1}^{n^{*}} x^{j}_{\{n\},\{a_{n}\}}(\sigma,\tau)
 \ee
$\sigma\in[0,1]$ and $\tau\in[0,\infty)$ are the worldsheet variables and $X^{j}(\sigma,\tau)$ is the map between the worldsheet and the target space (${\mathbb{R}}_{3}$), representing the 3-D string profile at any value of $\tau$\footnote{We did not include the center of mass position of the string $x_{0}$, and also the center of mass momentum $p_{0}\tau$, because are not relevant in our investigation. }.
Classical string configurations can be classified by the set of harmonics $\{n\}$ and the respective set of coefficients $\{a^{j}_{n}\}$ which are respectively the harmonic label and the relative weight coefficient of the classical mode $x^{j}$ :
\be\label{fbk2}
x^{j}_{\{n\},\{a_{n}\}}(\sigma,\tau)= {a_{n}^{j}\over n} \cos{(n\pi \sigma)} \sin{(n\pi \tau)}
\ee
In figure \ref{f1} there are some 3-D string profiles for different choices of the set of harmonics and coefficients, from which one can observe how the string profile becomes more involved if the number of harmonics is increased.

\begin{figure}[h!] 
\centering
\includegraphics[scale=0.35]{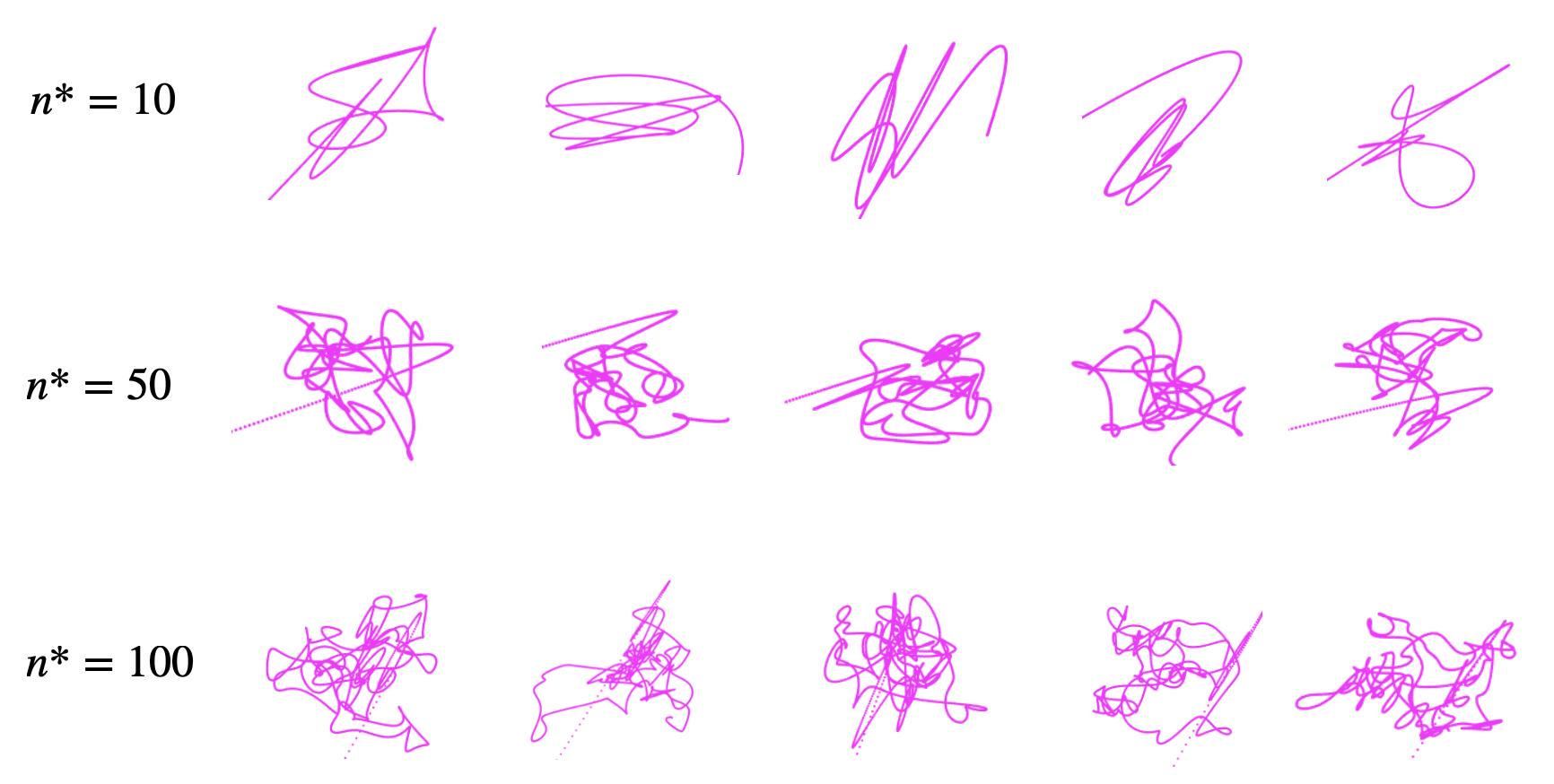}
\caption{Examples of 3D string profiles for different combinations of harmonics  $n^{*}=10,50,100$. For each value of $n^{*}$ there are five choices of string profiles for different choices of uniformly distributed random parameters $a_{n}^{j}\in (0,1)$. } \label{f2}
\end{figure}

From figure \ref{f2} one can observe the dependence of string profiles from the set of coefficients $\{a_{n}^{j}\}$. In the present case it was assumed that coefficients are normalized to the identity, and they are uniformly distributed in the interval $(0,1)$. One can observe that the behavior of  string profiles is unchanged  also for non normalized integer coefficients. The features of string profiles are unaffected by different parametrizations of the coefficients $a_{n}^{j}$.
At this level a trivial observation is that, if one considers a classical string profile
\be\label{fbk3}
 X^{j}_{\{a_{n}\}}(\sigma,\tau)=x^{j}_{\{n\},\{a_{n}\}}(\sigma,\tau)
 \ee
  with the same single generic harmonic $n$ in the three spatial directions $j=1,2,3$, one obtains the smoothest profile, which is a straight line, such as the first profile of figure \ref{f1}. 
  
This is the general picture of how the classical string profiles are modeled by the harmonic set $\{n\}$ and the coefficient set $\{a_{n}\}$. 

Now it is helpful to study the comparison between classical and quantum string configuration. In particular, promoting the coefficients $a_{n}^{j}$ to be creation operators ${\cal A}_{-n}^{j}$ one has the quantum analog of the string mode, and in addition to the set of harmonics $\{n\}$, one has to include the number of excitation $g_{n}$ of each harmonic so much so that for a given quantized level $N$ of the string spectrum, one has a set of states spanned by the set of solutions $\{g_{n}\}$, representing all the partitions of the integer $N$ with occupation number $J$:
\be\label{fbk4}
N=\sum_{n=1}^{N}n g_{n}\,,\quad J=\sum_{n=1}^{N}g_{n}
\ee
The promotion from classical to quantum string is summarized as
\be\label{classTOquant}
x_{\{n\},\{a_{n}\}}(\sigma,\tau)\,\Rightarrow \, {\cal N}_{n,g_{n}} {\cal A}_{-n}^{g_{n}} | 0\rangle\,\,; \quad  X_{\{a_{n}\}}(\sigma,\tau)\Rightarrow \Pi^{(N)}_{\{g_{n}\}}= \prod_{n=1}^{N} {\cal N}_{n,g_{n}} {\cal A}_{-n}^{g_{n}} | 0\rangle
\ee
where ${\cal N}^{-1}_{n,g_{n}}=\sqrt{n^{g_{n}}\,g_{n}!}$ is the normalization constant of each mode\footnote{For simplicity we have suppressed the space time index $j$}.
In particular due to its quantum nature, the quantum string configuration for a fixed level $N$ has a large degeneracy\footnote{The degeneracy of states at large $N$ grows like $\sim e^{\sqrt{N}}$} which produces the microstate structure, so in order to identify the final quantum string state one can take the simplest linear combination of microstates introducing the average over microstates in the second expression of (\ref{classTOquant})
\be\label{fbk5}
X(\sigma,\tau)\Rightarrow \sum_{\{g_{n}\}} \Pi^{(N)}_{\{g_{n}\}}=\sum_{\{g_{n}\}} \prod_{n=1}^{N} {\cal N}_{n,g_{n}}{\cal A}_{-n}^{g_{n}} | 0\rangle
\ee
Looking at the complicated classical string profiles, parametrized by $\{n\}$ and $\{a_{n}\}$, a natural question is how the quantum string configurations reflect their classical characteristic shape as a function of $\{n\}$ and $\{g_{n}\}$.

A possible way of testing the features of quantum string profiles as a function of the microstate structure, is to probe the implications of their shapes through their interactions. In particular one can choose the simplest string decay amplitude and study the microstate dependence.

In the next subsection there will be a review of the analysis of string configurations leading to a chaotic behavior of their interactions \cite{ChaosScatt}.

\subsection{Probing chaotic behavior of quantum strings through their interactions}
\begin{figure}[h!]
\centering
\includegraphics[scale=0.35]{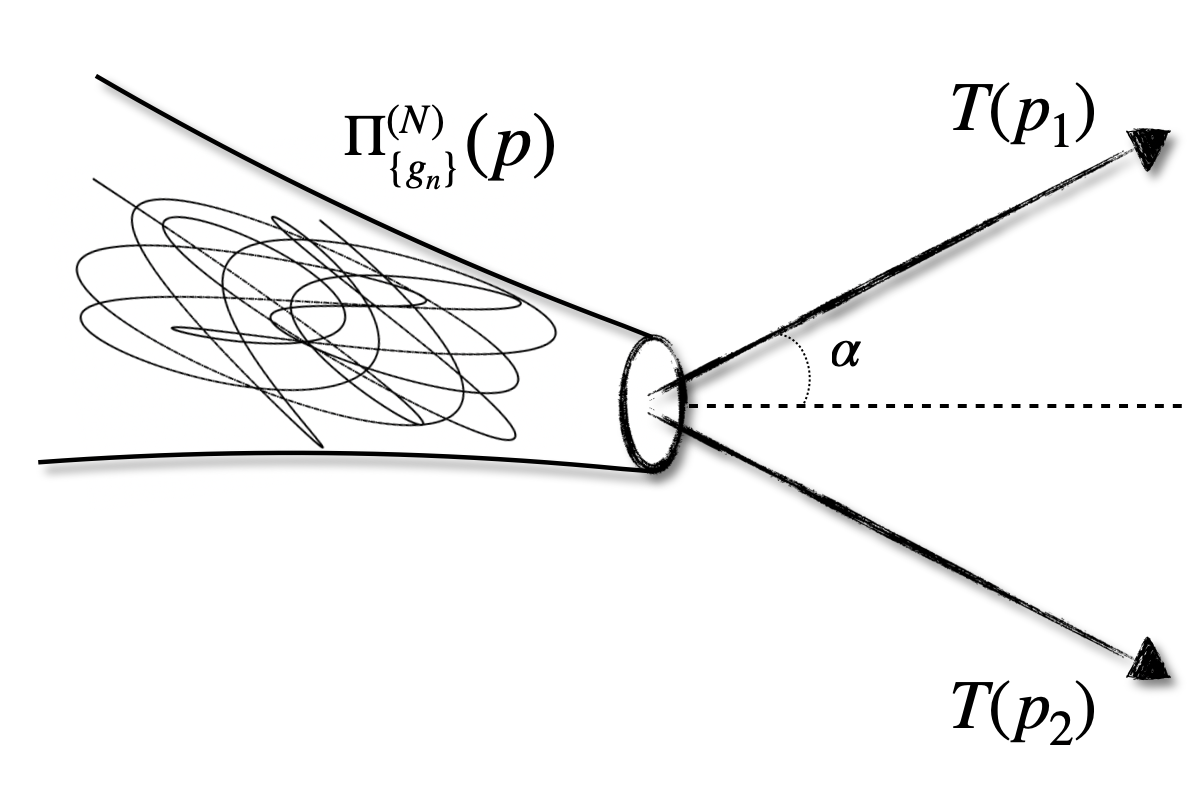}
\caption{Picture of the decay amplitude where a representative microstate $\Pi_{\{g_{n}\}}^{(N)}$ decays into two tachyons $T$. The only kinematical freedom of the decay amplitude is the emission angle $\alpha$ which starts from the reference dashed line.} 
\label{figScattHTT}
\end{figure}

Inspired by the systematic of classical string profiles, the logic proposed for the analysis of quantum string configurations is the following: the main observable is the decay amplitude provided by a level $N$ string microstate $\Pi_{\{g_{n}\}}^{(N)}$ decaying into two tachyons. Along the line of indications and improvements introduced in \cite{Skliros:2011si} and developed in \cite{Bianchi:2019ywd}, the decay amplitude ${\cal A}_{\Pi^{(N)}_{\{g_{n}\}}}$ for the most general HES was computed, with a remarkably simple procedure based on coherent state techniques. The informations about the structure of $\Pi_{\{g_{n}\}}^{(N)}$ are translated into the decay amplitude, and they are manifested through the profile of the decay amplitude. The choice of looking at tachyons is connected to their simple vertex operators, in such a way that all the information inside the decay amplitude is governed by the structure of $\Pi_{\{g_{n}\}}^{(N)}$. In figure \ref{figScattHTT} there is a representative picture of the decay amplitude from which one can see that the decay amplitude profile is only a function of the angle $\alpha$. Now comparing decay profiles of different microstates one can extract a general behavior associated to choices of $\{n\}$ and $\{g_{n}\}$, as it was pointed out in \cite{Gross:2021gsj}. Most of the information of the decay profile can be codified in terms of its extrema, so it is useful to introduce the logarithmic derivative $F_{\{g_{n}\}}(\alpha)$ and study its distribution of zeros. A suitable parameterization of the distribution is intimately connected with the chaotic behavior of the decay \cite{ChaosScatt}. Before describing the chaotic analysis in detail, a fast presentation of the resulting decay amplitude is discussed.

The HES state of the level $N$ with polarizations $\{\zeta_{n}\}$ and momentum $p$ is described by
\be\label{fbk6}
\Pi_{\{g_{n}\}}^{(N)}(\{\zeta_{n}\},p)=\prod_{n=1}^{N}{\left(\zeta_{n}{\cdot}{\cal A}_{-n}\right)^{g_{n}}\over \sqrt{g_{n}! \,n^{g_{n}}}} | \widetilde{p}\rangle
\ee
where $\widetilde{p}$ is the tachyonic DDF reference momentum, that combined with the action of the creation operators reproduce the momentum of the final state $p=\widetilde{p}-q\sum_{n}ng_{n}$. Following the DDF formalism one can write the corresponding BRST vertex operator and compute the decay amplitude of figure \ref{figScattHTT} using circular polarizations\footnote{The general amplitude is made of contributions both linear and bilinear in $\zeta_{n}$ \cite{Bianchi:2019ywd}. The bilinear contribution is proportional to the square of the linear contribution, so up to an irrelevant polynomial the functional dependence on the microstate structure is preserved by the choice $\zeta^{(a)}_{n}{\cdot}\zeta^{(b)}_{m}=0$.}:
\be\label{fbk7}
{\cal A}_{\Pi^{(N)}_{\{g_{n}\}}}(\alpha)\simeq \Big(\zeta{\cdot}(p_{1}{-}p_{2}) \Big)^{J}\prod_{n=1}^{N} \left( { (1{-}n \sin^{2}\alpha)_{n{-}1}  \over \Gamma(n)}\right)^{g_{n}}
\ee
all the information about the microstate $\Pi_{\{g_{n}\}}^{(N)}$ is encoded in the dressing factor of the coupling $\zeta{\cdot}(p_{1}{-}p_{2})$
\be\label{chaosHTT}
\Pi_{\{g_{n}\}}^{(N)}{-}\text{structure}\Rightarrow \prod_{n=1}^{N} \left( { (1{-}n \sin^{2}\alpha)_{n{-}1}  \over \Gamma(n)}\right)^{g_{n}}
\ee
Using the properties of the Pochhammer factor and the explicit parametrization of the polarizations the decay amplitude can be written as 
\be\label{fbk8}
{\cal A}_{\Pi^{(N)}_{\{g_{n}\}}}(\alpha)\simeq \prod_{n=1}^{N}\left({\sin{\alpha}\over \Gamma(n)}\, \sin{\left(n\pi \cos^{2}\alpha/2 \right)} \, \Gamma\left( n\cos^{2}\alpha/2 \right)\,\Gamma\left( n\sin^{2}\alpha/2 \right) \right)^{g_{n}}
\ee
Finally the logarithmic derivative of the decay amplitude
\be\label{fbk9}
F_{\{g_{n}\}}(\alpha)={d\over d\alpha}\log{{\cal A}_{\Pi^{(N)}_{\{g_{n}\}}}(\alpha)}
\ee
 has the following form
\be\label{logder}
\begin{split}
F_{\{g_{n}\}}(\alpha)=&J \cot\alpha- \pi{\sin\alpha\over 2}\sum_{n=1}^{N} n g_{n} \cot{\left( n\pi \cos^{2}{\alpha\over2} \right)}\\
& {-}{\sin\alpha\over 2} \sum_{n=1}^{N}n g_{n} \Big( \psi\left( n \cos^{2}{\alpha\over2} \right){-}\psi\left( n \sin^{2}{\alpha\over 2} \right) \Big).
\end{split}
\ee
This is the final observable which will be subjected to the analysis described below.

\subsubsection*{Chaotic analysis: setup}
Random Matrix Theory (RMT) provides a very powerful tool to make the bridge between quantum chaos and universal statistical properties \cite{RMTbook}\cite{ChaosNucl}. In particular a quantitative connection between chaos and probability distributions was conjectured in \cite{ChaosStat} and subsequently the link between chaos and statistical properties was widely studied in many contexts such as quantum chromodynamics (QCD)\cite{Verbaarschot:2000dy}, nuclear physics \cite{Weidenmuller:2008vb}, black holes \cite{Cotler:2016fpe} and condensed matter \cite{Guhr:1997ve}. In what follows we laid out the identification strategy of the target distribution used as a discriminant of the chaotic behavior.

\begin{itemize}
\item Starting from the Hermite $\beta$-ensemble of $N\times N$ random matrices, given the set $\{\alpha\}=(\alpha_{1},....,\alpha_{N})$ of matrix eigenvalues, the associated joint probability distribution is given by
\be\label{fbk10}
P_{\beta}(\{\alpha\})={e^{-\sum_{j=1}^{N} {\alpha_{j}^{2}\over 2} }\over Z_{N,\beta}} \prod_{\ell< v} |\alpha_{\ell}-\alpha_{v}|^{\beta}
\ee
\item A very useful approximation was given in \cite{WigSurm}, where it was considered the joint probability distribution of nearest-neighbor spacing $\alpha_{j+1}-\alpha_{j}=\delta_{j}$
: the Wigner surmise distribution 
\be\label{fbk11}
P_{\beta}(\{\delta\})=C_{\beta}\, \delta^{\beta}\,e^{-d_{\beta}\delta^{2}}
\ee
with constants
\be\label{fbk12}
C_{\beta}=2 \, {\Gamma(\beta/2+1)^{\beta+1}\over \Gamma(\beta/2+1/2)^{\beta+2} }\,, \quad d_{\beta}={\Gamma(\beta/2 + 1)^{2}\over \Gamma(\beta/2 + 1/2)^{2} }
\ee
A very nice application of the Wigner surmise is the prediction of the zeros distribution of the Riemann zeta function \cite{Rzeros}.

\item In considering the joint probability distribution of consecutive spacings, there is a technical  issue related to the unfolding procedure of the data, that in general cases can be difficult to implement, as explained in \cite{rIndex}. In order to avoid the unfolding procedure one can introduce a more robust index \cite{rIndex}: the ratio of consecutive level spacing
\be\label{rind}
 r_{j}={\delta_{j+1}\over \delta_{j} }
 \ee
 where the joint probability distribution was intensively studied in \cite{rIndexAna3}\cite{rIndexAna4}, and for $3 \times 3$ block diagonal matrices takes the following form
\be\label{pdfRindex}
P_{\beta}(r)= {3^{3(1+\beta)/2} \Gamma(1+\beta/2)^{2}\over 2\pi \Gamma(1+\beta) }\, {(r+r^{2})^{\beta}\over (1+r+r^{2})^{1+3\beta/2}}
\ee
which is valid for any value of $\beta>0$ \cite{rIndexAnaExt} and also for large asymptotic values of $\beta$ \cite{rIndexAnaAsym}. For $\beta=1,2,4$ one has the standard GOE, GUE and GSE respectively, which are the gaussian orthogonal/unitary/symplectic ensembles. 
\end{itemize}

\subsubsection*{Chaotic analysis: results}
\begin{figure}[h!]
\centering
\includegraphics[scale=0.55]{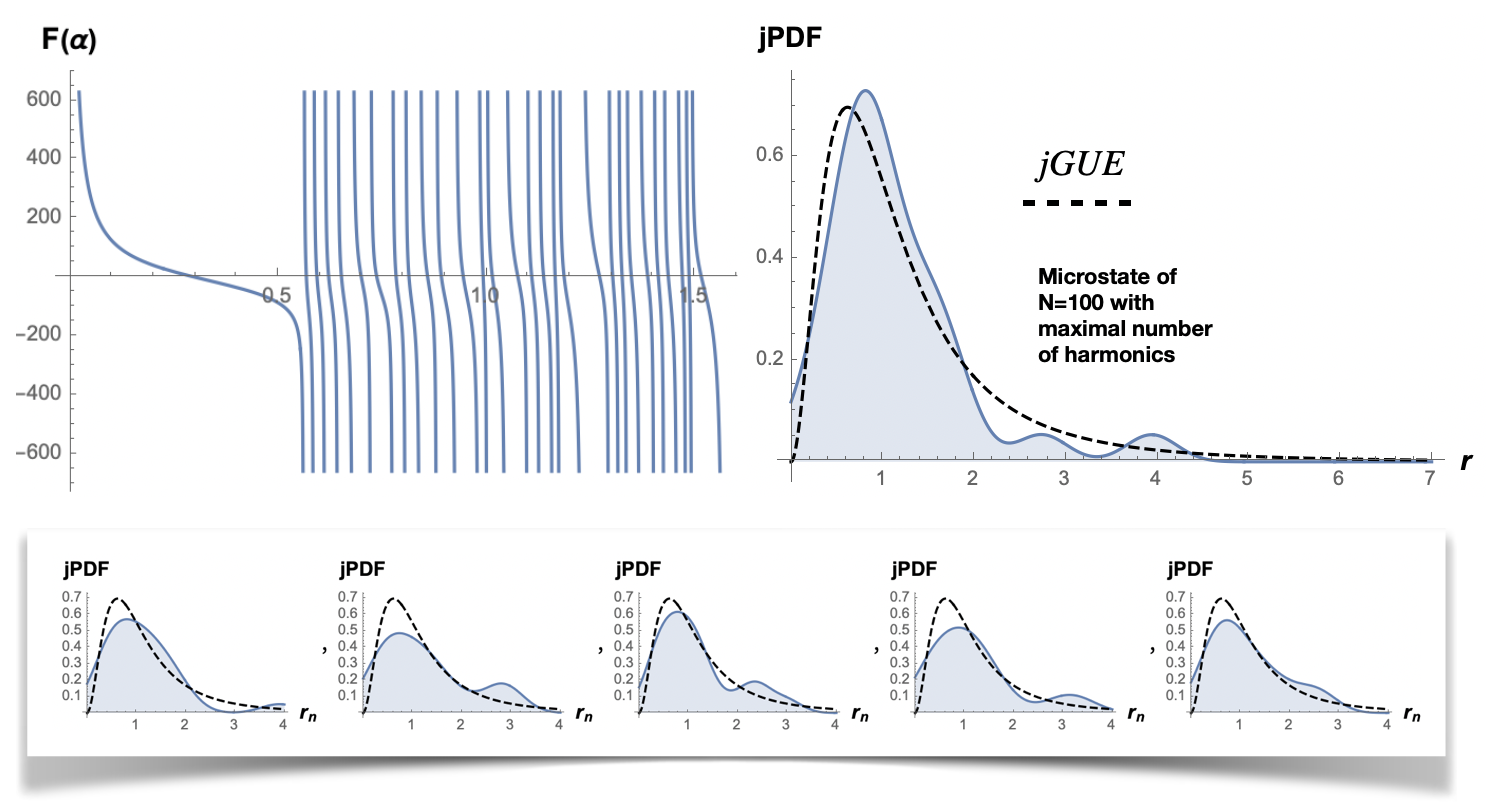}
\caption{Microstates of $N=100$. From left to right the profile of the logarithmic derivative of the decay amplitude and the joint probability distribution both relative to the microstate with maximal number of harmonics. The dashed line is the joint probability distribution with $\beta=2$, which represents the joint GUE distribution (jGUE). Below there are five examples of joint probability distributions for microstates with less number of harmonics.}
\label{figN100}
\end{figure}

\begin{figure}[h!]
\centering
\includegraphics[scale=0.35]{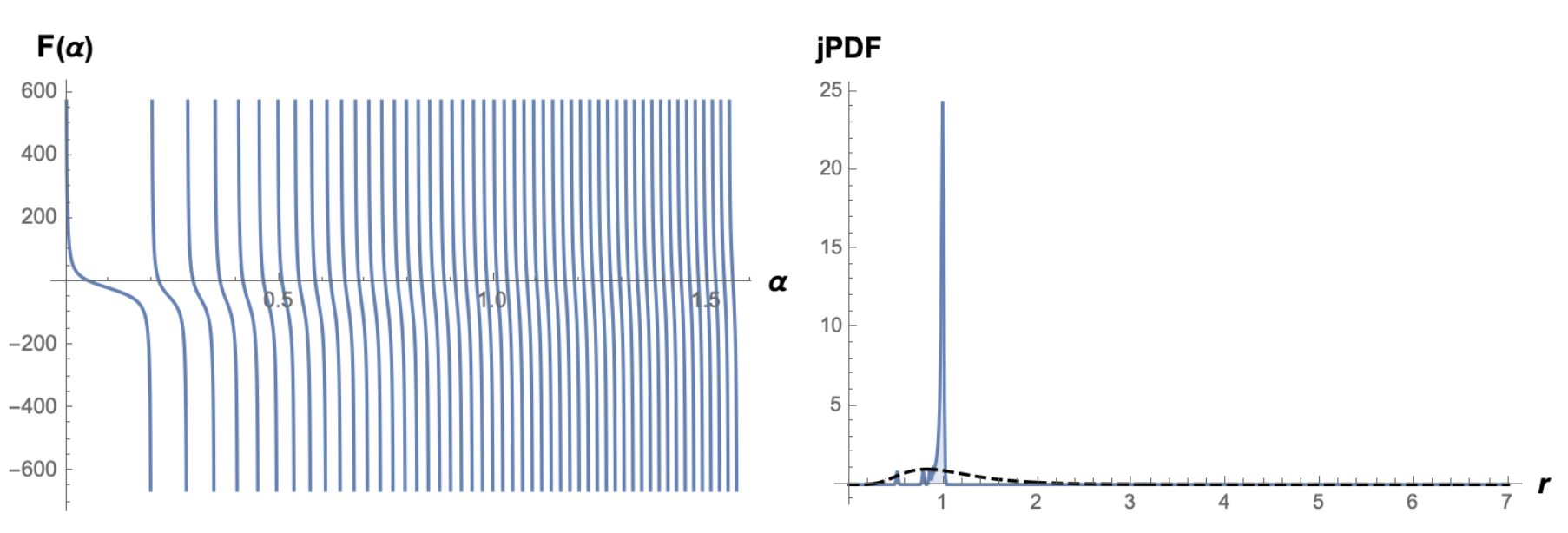}
\caption{The logarithmic derivative of the decay amplitude and the joint probability distribution both relative to the microstate of $N=100$ with only one harmonic. The dashed line is the jGUE.}
\label{trivfigN100}
\end{figure}

Following the discussion related to figure \ref{f1} one can expect that the microstate with the maximal number of harmonics will produce a less smooth decay amplitude. In particular one can measure the chaotic behavior of the decay profile computing the distribution of the index (\ref{rind}) for the zeros of (\ref{logder}), which is the study of how the unbiased $r$-index indicator(\ref{rind}) for the extrema of the amplitude is distributed in agreement with the target chaotic class of $\beta$-distributions (\ref{pdfRindex}). From the results in figure \ref{figN100}, relative to microstates of $N=100$, one can observe the profile of the logarithmic derivative $F_{\{g_{n}\}}(\alpha)$ and the respective joint probability distribution of the microstate with the maximal number of harmonics (which are the first two plots respectively). The five small plots represent how the measured joint probability distributions deviate from the target distribution (\ref{pdfRindex}), respect the variation of the number of harmonics. Quite similar to the case of classical string profiles where the number of harmonics triggers the complexity of shape of the profiles, the chaotic behavior of the decay profile is triggered by the number of harmonics. An additional check of this kind of harmonic hierarchy is provided in figure \ref{trivfigN100}, where we plot the logarithmic derivative of the decay amplitude and the joint probability distribution of the single harmonic microstate of $N=100$. In particular one can observe that the distribution totally deviates from the chaotic jGUE, which is a hint about the connection between classical single harmonic string profiles (straight lines) and the absence of chaos in the decay amplitude of single harmonic microstates.
\begin{figure}
\centering
\includegraphics[scale=0.65]{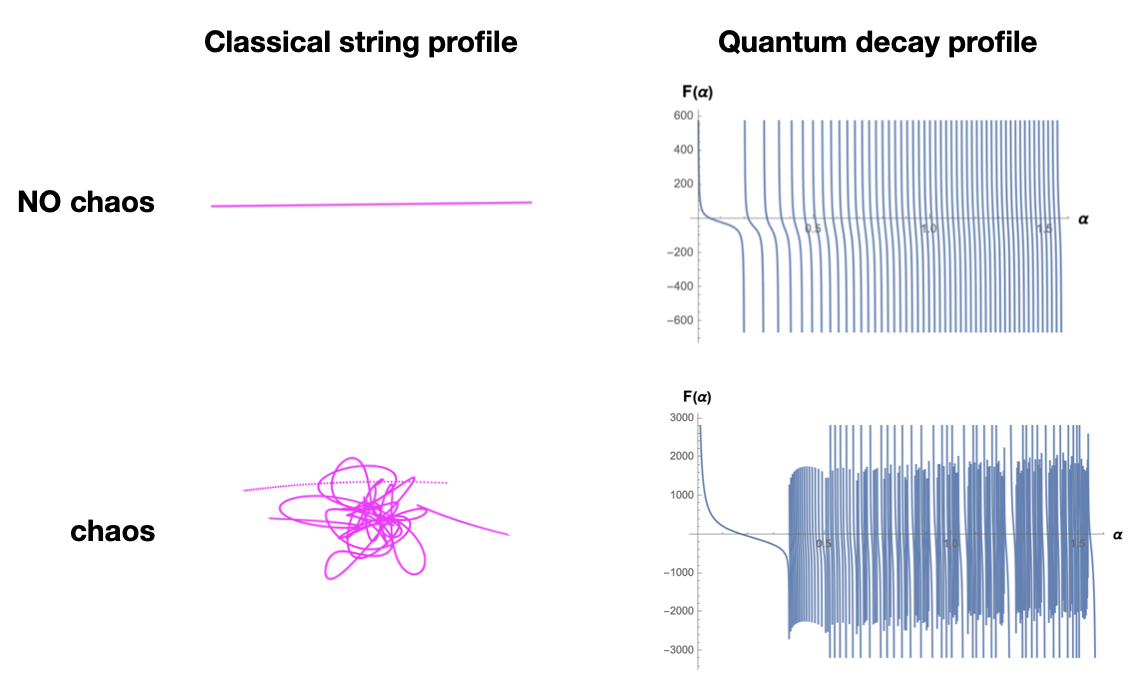}
\caption{Comparison between classical string profiles and decay amplitudes}
\label{claVSqua}
\end{figure}

\newpage
\section{Thermalization emergence in highly excited string decays}\label{se2}
\begin{figure}[h!]
\centering
\includegraphics[scale=0.45]{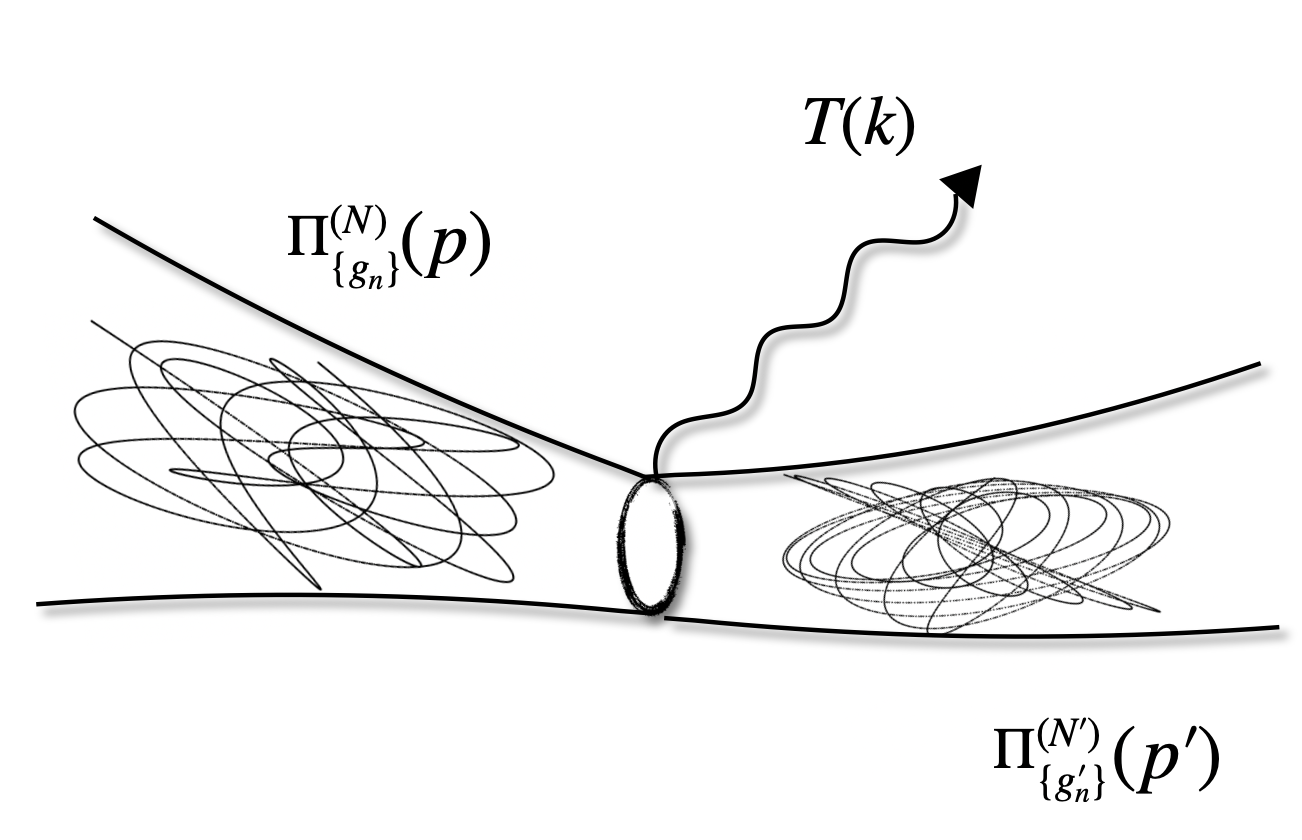}
\caption{Picture of the decay amplitude where a representative microstate $\Pi_{\{g_{n}\}}^{(N)}$ decays into $\Pi_{\{g'_{n'}\}}^{(N')}$ emitting a tachyon $T$.} 
\label{fiScattHHT}
\end{figure}

In the previous section we presented a systematic study of how the microstate structure emerges through the profile of its decay process. In particular we described how the chaotic behavior of the decay process is triggered by the microstate structure. The aim of this section is to study how the chaotic behavior is related to the thermalization process of the most general HES decay (fig.\ref{fiScattHHT}). 

Following \cite{Firrotta:2022cku} and appendix (\ref{appA}) the decay rate of the present process, in a d-dimensional phase space, is given by the following expression
\be\label{fbk13}
{\Gamma}_{\Pi^{N}_{\{g_{n}\}},\Pi^{N'}_{\{g_{n'}\}}}= {\Omega_{s}\,E_{k}^{d-3}\over 16 (N{-}1)(2\pi)^{d-2}}{\Big|{\cal A}_{\Pi^{N}_{\{g_{n}\}},\Pi^{N'}_{\{g_{n'}\}}}\Big|^{2}}
\ee
This is the decay rate of the single microstate  at level $N$ decaying into a single microstate at level $N'$ through the emission of a tachyon of energy $E_{k}$, where the solid angle $\Omega_{s}=2\pi^{(d{-}1)2}/\Gamma((d{-}1)/2)$ is introduced.

The non trivial dependence of the decay rate is due to the square of the absolute value of the amplitude, which is the main quantity that will be analyzed. In particular in the region where the ratio between the energy of the emitted state and the mass of the decaying state is enough small, in such a way that the energy loss of the decaying state is smooth, the absolute value square of the amplitude assumes the following form
\be\label{thermA}
{\Big|{\cal A}_{\Pi^{N}_{\{g_{n}\}},\Pi^{N'}_{\{g_{n'}\}}}\Big|^{2}}= \widetilde{{\cal N}}_{\{g_{n}\}}^{2} \widetilde{{\cal N}'}_{\{g'_{n'}\}}^{2} \, e^{-{\cal C}_{N}(\{g_{n}\},\{g'_{n'}\}){E_{k}\over T_{H}}-2\mu_{N}\left( \{g_{n}\},\{g'_{n'}\};E_{k}/T_{H}\right)}
\ee
where there is a weight factor ${\cal C}_{N}$ that depends on the level $N$ of the decaying state, and also depends on the microstates geometry through the relation
\be\label{fbk14}
{\cal C}_{N}(\{g_{n}\},\{g'_{n'}\})={2\over \sqrt{N}}\left(\sum_{n=1}^{N}g_{n} n\log{n} - \sum_{n'=1}^{N'}g'_{n'} n' \log{n'}\right)
\ee
the other function in the exponent is given by
\be\label{fbk15}
\mu_{N}\left( \{g_{n}\},\{g'_{n'}\};{E_{k}\over T_{H}}\right)= \sum_{n=1}^{N} g_{n} \log\Gamma\left( 1{-}{n\over \sqrt{N}} {E_{k}\over T_{H}}\right) +  \sum_{n'=1}^{N'} g'_{n'} \log\Gamma\left( 1{+}{n'\over \sqrt{N}} {E_{k}\over T_{H}}\right).
\ee
and finally the dimensional temperature $T_{H}=1/ \ell_{s}$ is the Hagedorn temperature which is the inverse of the string length $\ell_{s}=\sqrt{\alpha'}$. 

The thermal nature of the characteristic expression (\ref{thermA}) is intimately related to the chaotic behavior of the decay, in fact the non trivial dependence on the microstates is directly related to chaos (section \ref{se1}) and it will play also a crucial role in the thermalization of the decay, as will be explained in the next part \ref{chaostherm} of the present section.

\begin{figure}[h!]
\centering
\includegraphics[scale=0.50]{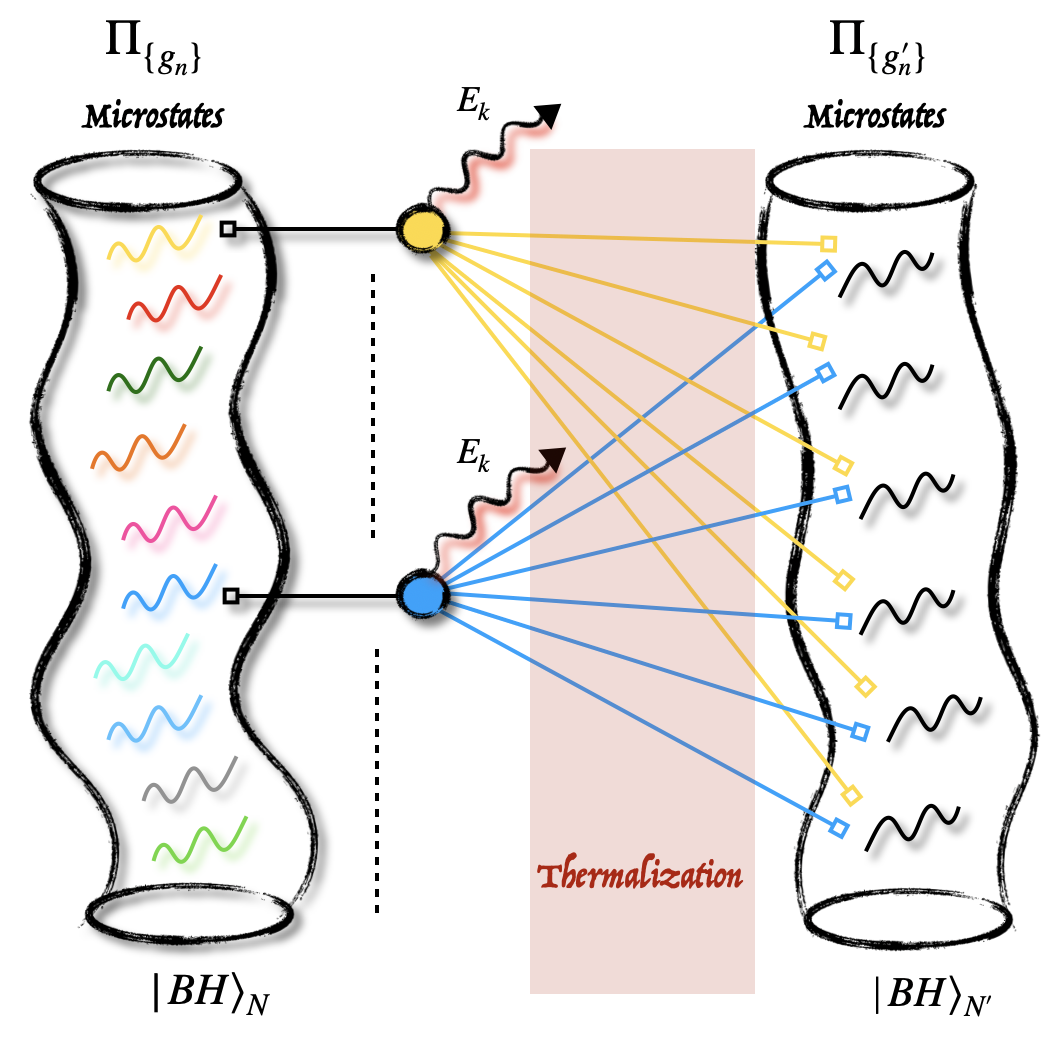}
\caption{Qualitative picture of the decay rate $|BH\rangle_{N}\Rightarrow |BH\rangle_{N'} +E_{k}$. The microstates of the decaying state are represented by colorful small strings, each color is associated to the possible microstate decay. The red region represents the random superposition of all the possible decays which is the mechanism through which the thermalization process emerge.}\label{thermopic}
\end{figure}

When is considered the decay rate of a state of the level $N$ made of many microstates (see fig.\ref{thermopic})
\be\label{fbk16}
|BH\rangle_{N}=\sum_{\{g_{n}\}} \Pi^{(N)}_{\{g_{n}\}}
\ee
which can be interpreted as a black hole state, the final decay rate results to be the sum over all the possible microstate configurations of the decay rates
\be\label{decayBH}
{\Gamma}\Big({|BH\rangle_{N}\Rightarrow |BH\rangle_{N'} +E_{k}}\Big)={\Omega_{s}\,E_{k}^{d-3}\over 16 (N{-}1)(2\pi)^{d-2}}\sum_{\{g_{n}\},\{g'_{n'}\}}{\Big|{\cal A}_{\Pi^{N}_{\{g_{n}\}},\Pi^{N'}_{\{g_{n'}\}}}\Big|^{2}}.
\ee
This is a very rich observable that is characterized by the highly non trivial functional dependence of the microstates. In the last part \ref{part2therm} of the present section, we will describe the behavior of such observable and the derivation of the non trivial energy spectrum leading to the greybody radiation of highly excited string states.

\subsection{Chaos driven thermalization}\label{chaostherm}
In section \ref{se1} we analyzed the chaotic behavior of a string microstate decaying into two tachyons, where we described the connection between the microstate structure and the chaotic behavior of the decay. The sensitivity of the decay to the microstate structure was essentially encoded in the dressing factor (\ref{chaosHTT}). Now considering the direct extension of the decay in (fig.\ref{figScattHTT}), which is the one of (fig.\ref{fiScattHHT}), one has the generalization of (\ref{chaosHTT}) to the case of two microstate structures: one for the decaying microstate $\Pi^{N}_{\{g_{n}\}}$ and the other for the final microstate $\Pi^{N'}_{\{g'_{n'}\}}$. Using the general setup described in appendix \ref{appA} one finds a remarkably compact expression of the decay rate\footnote{In the expression we used $\alpha'=1/2$ that means $T_{H}=\sqrt{2}$.}
\be\label{dHHT}
{\Big|{\cal A}_{\Pi^{N}_{\{g_{n}\}},\Pi^{N'}_{\{g_{n'}\}}}\Big|^{2}}\simeq\hspace{-1 mm} \prod_{n=1}^{N}\left({\left(1{-}n(1{-} E_{k}/\sqrt{2N}) \right)_{n{-}1}\over \Gamma(n)}\right)^{2g_{n}}\hspace{-3 mm}\prod_{n'=1}^{N'}\left({\left(1{-}n'(1{+}E_{k}/\sqrt{2N}) \right)_{n'{-}1}\over \Gamma(n')}\right)^{2g'_{n'}}
\ee
comparing this expression with (\ref{chaosHTT}) one can see the systematic generalization of the decay rate due to the presence of an additional microstate. Following the expansion (\ref{exptherm}) and the rest of the appendix, one recovers (\ref{thermA}). This is the systematics of how the chaotic factors make the decay rate thermal, giving a precise identification between chaos and the emergence of thermalization.

Following the same logic as in (fig.\ref{claVSqua}), where a comparison between classical strings and decay amplitudes was argued, one can complete the scenario describing the mechanism that originates the thermalization of the precess. 

In fact one can analyze two extreme cases: the first one is to consider microstates of the first Regge trajectory. Classically their profile is just a straight line (fig.\ref{claVSqua}), because of the fact that they are single harmonic states, so they do not produce chaos. The associated thermalization is given by the expression (\ref{dHHT}) with $n=1,\,g_{1}=N$ and $n'=1,\,g'_{1}=N'$, which trivializes to unity, so the decay is not thermal.  

The second extreme case, which is less trivial, is to consider still single harmonic states but with the maximal harmonic excited, $n=N,\,g_{N}=1$ and $n'=N',\,g'_{N'}=1$. Classically they are a straight lines and do not produce chaos (fig.\ref{trivfigN100}). From (\ref{dHHT}) and (\ref{thermA}) one can finds
\be\label{fbk17}
{\Big|{\cal A}_{\Pi^{N}_{\{g_{N}=1\}},\Pi^{N'}_{\{g_{N'}=1\}}}\Big|^{2}}\simeq e^{-2\log\left({N\over N'}\right)\sqrt{N}E_{k}/T_{H}} \left({\sin\left({\pi \sqrt{N}E_{k}/T_{H}}\right)\over \pi \sqrt{N}E_{k}/T_{H}}\right)^{2}
\ee
this behavior, at large $N$, is very suppressed even if $N'\sim N$, and it reflects how the thermal behavior of the decay is subdominant in the case of states with only the maximal harmonic excited.  

\begin{figure}[h!]
\centering
\includegraphics[scale=0.5]{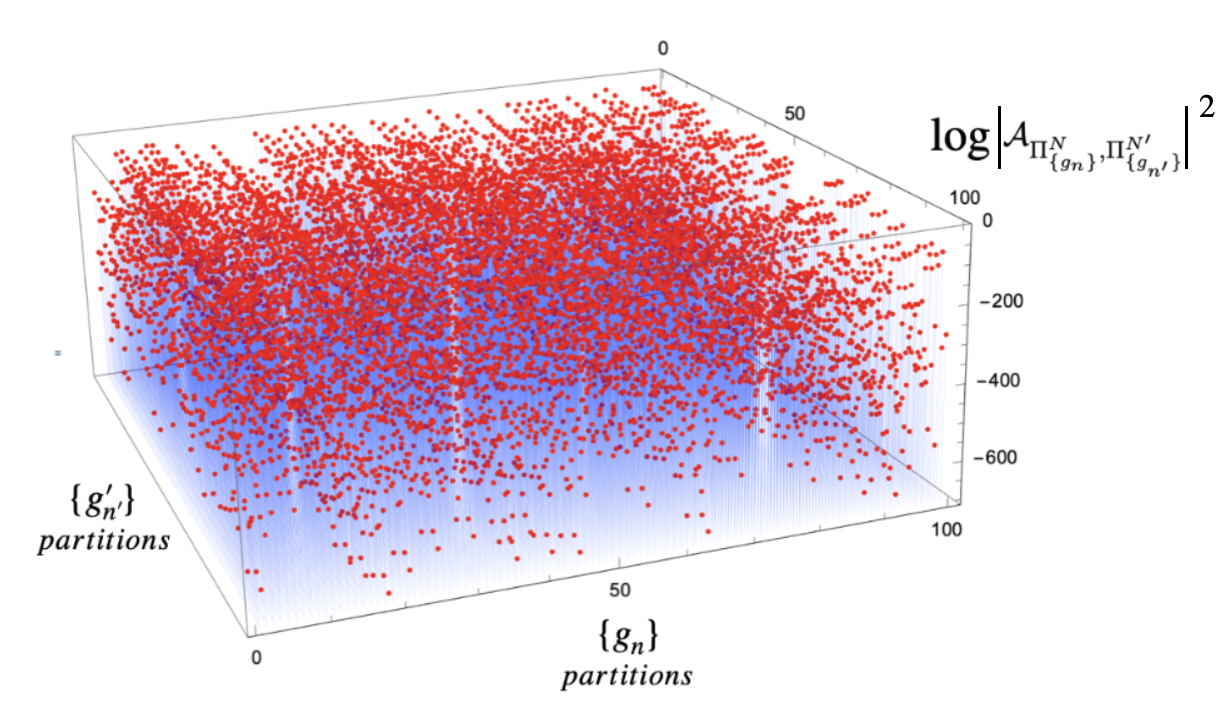}
\caption{Microstates population of the decay rate in logarithmic scale for 100 random partitions $\{g_{n}\}$ of $N=100$ and 100 random partitions $\{g'_{n'}\}$ of $N'=99$. The red points are the values of the decay rate.}\label{distmicr100}
\end{figure}

\begin{figure}[h!]
\centering
\includegraphics[scale=0.5]{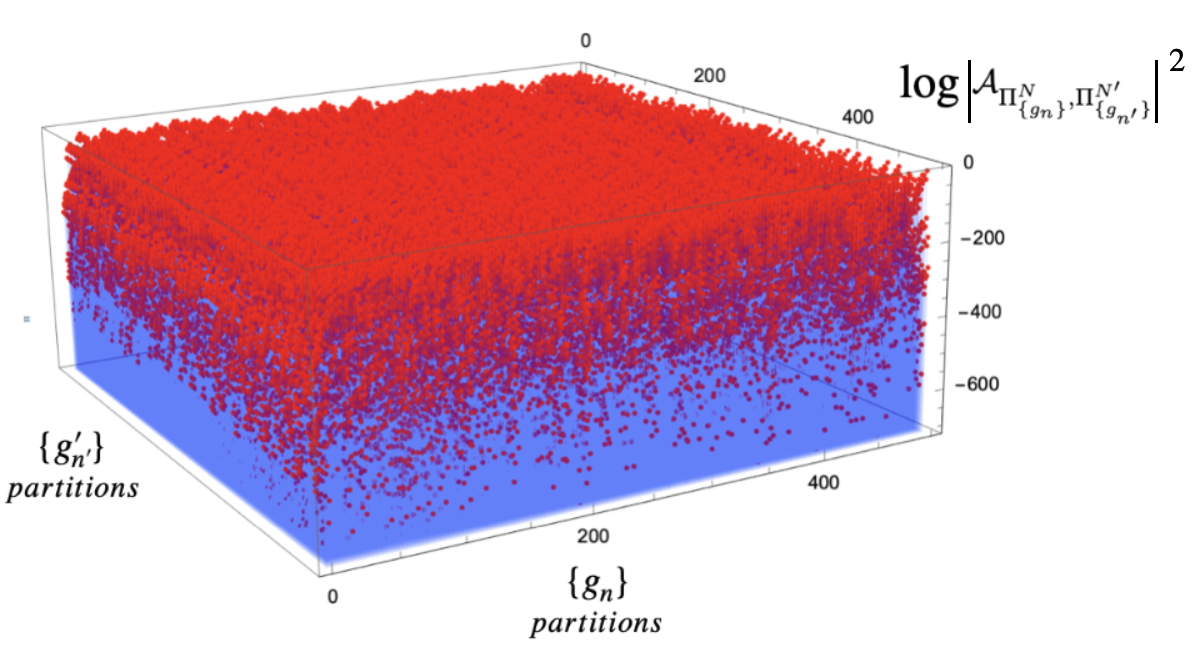}
\caption{Microstates population of the decay rate in logarithmic scale for 500 random partitions $\{g_{n}\}$ of $N=100$ and 500 random partitions $\{g'_{n'}\}$ of $N'=99$. The red points are the values of the decay rate.}\label{distmicr500}
\end{figure}

Beyond the two extremal cases discussed, one can have a more qualitative picture of the microstates functional dependence of the decay looking at the explicit distributions in (fig.\ref{distmicr100}) and (fig.\ref{distmicr500}). 

 The fluctuations of the decay reflect the connection between the geometry of microstates, which is clear in the classical picture of string profiles, and the associated chaos which is the trigger of the thermal behavior. 

\subsection{Thermal spectrum: the greybody emission of highly excited strings}\label{part2therm}
\begin{figure}[h!]
\centering
\includegraphics[scale=0.6]{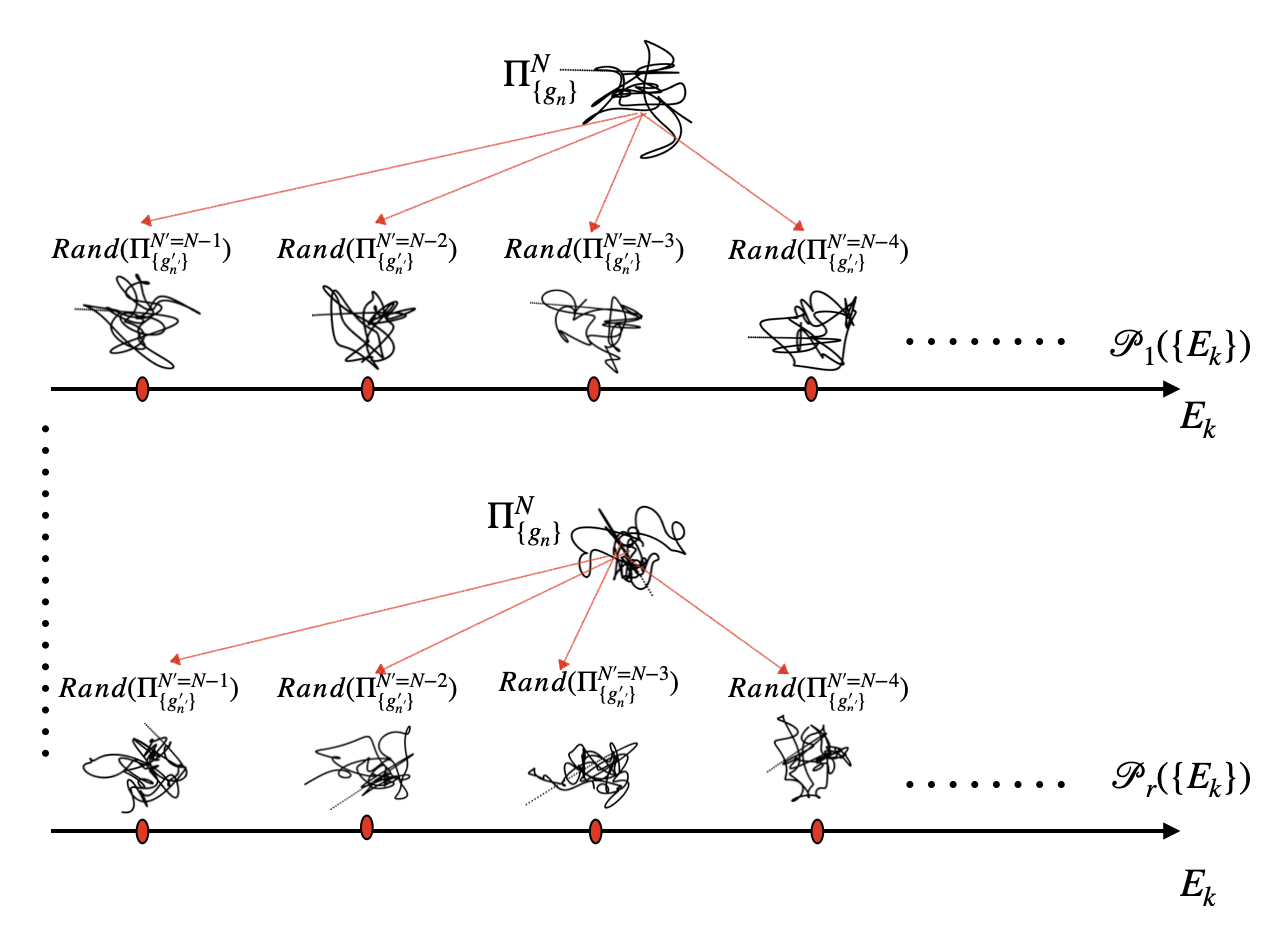}
\caption{Picture of the random microstates energy trajectories ${\cal P}(\{E_{k}\})$.}\label{thermspec}
\end{figure}

In the previous part we described the connection between chaos and thermal behavior of the decay rate, where the intrinsic microscopical structure of microstates was fundamental for the origin of such connection. Here the goal will be the analysis of the energy spectrum radiated from HES states. The first thing to note is that the decay rate (\ref{thermA}) is not a function of the emitted energy $E_{k}$, the kinematics of the two body decay fixes the energy $E_{k}$ as a function of the masses of the present states
\be\label{fbk18}
E_{k}={M_{N}^{2}-M_{N'}^{2}+M_{T}^{2}\over 2 M_{N}}={N-N'-1\over \sqrt{2N-2}}
\ee
In order to extract the energy spectrum of an HES state, one has to explore the energy range of the decay rate point by point, producing a discrete energy trajectory of the decay process. Since the target observable is the radiation of the HES state at generic level $N$, one can reproduce the energy trajectory varying the mass $M_{N'}=2N'-2$ of the final state. The energy region of interest is identified by the range in which the ratio between the energy of the emitted state $E_{k}$ and the mass of the decaying state $M_{N}=2N-2$ is enough small in such a way that the energy loss of the decaying state is smooth, which is the situation where a thermal spectrum is expected.
Given $N$ and $N'$, the decay rate of a state made of microstates is given in (\ref{decayBH}), and even if the energy $E_{k}$ is fixed there is a non trivial microstates distribution that mediates the decay rate (fig.\ref{distmicr500}). Therefore in order to reproduce the energy spectrum of a given HES state with ${\small\big\{\Pi^{(N)}_{\{g_{n}\}}\big\}}$ microstates, one can generate an energy trajectory of random decays for each microstate ${\small\Pi^{(N)}_{\{g_{n}\}}}$ of the level N that can decay into a random microstate ${\small\Pi^{(N')}_{\{g'_{n'}\}}}$ of the whole set ${\small\big\{\Pi^{(N')}_{\{g'_{n'}\}}\big\}}$ of the level $N'$. Finally the spectrum is obtained averaging over the energy trajectories (fig.\ref{thermspec}). In general each microstate of the level $N$ can decay into each microstate of the level $N'$, and this is the reason why one has to mimic the randomness of the decay process in order to describe the intrinsic nature of a degenerate sized object such as the HES. 

The general behavior of the energy spectrum turns out to be described by the greybody radiation
\be\label{gbspect}
\begin{split}
\Sigma^{\small{\text{grey}}}_{N}(E_{k}/T_{eff})&=\langle{\Gamma_{N\Rightarrow N'}}\rangle_{{\cal P}(\{E_{k}\})}={\sigma^{\text{grey}}_{N}(E_{k}/T_{eff})\over e^{{E_{k}\over T_{eff}}}-1}
\end{split}
\ee
with an effective temperature which depends on the decaying state through
\be\label{fbk19}
T_{eff}=T_{H}/\sqrt{N}
\ee
and with the greybody factor which is sensitive to the nature of the microstates of $N$ and $N'$ and also depends on the randomness of the decay process  
\be\label{fbk20}
\sigma^{\text{grey}}_{N}(E_{k}/T_{eff})= {C_{N}(\{g_{n}\}.\{g'_{n'}\})\Omega_{s}E_{k}^{d-3}\over 16 (N{-}1)(2\pi)^{d-2}}{\Big(e^{{E_{k}\over T_{eff}}}-1\Big) \left({E_{k}\over T_{eff}}\right)^{r_{N}(\{g_{n}\},\{g'_{n'}\})}\over e^{\nu_{N}(\{g_{n}\},\{g'_{n'}\}){E_{k}\over T_{eff}}}-1}
\ee
The randomness of the decay process is incorporated in the coefficients 
\be\label{fbk21}
C_{N}(\{g_{n}\},\{g'_{n'}\})\,,\quad r_{N}(\{g_{n}\},\{g'_{n'}\})\,,\quad \nu_{N}(\{g_{n}\},\{g'_{n'}\})
\ee
that also reflect the intrinsic dependence on the chosen microstates that determine the decaying state and the final state.

In the next section we will perform the spectrum analysis of different states, in particular a state with definite mass and occupation number $(N,J)$ decaying into a generic state of the level $N'$, a state $(N,J)$ decaying into a state $(N',J')$ and finally a generic state $N$ decaying into a state $N'$. For simplicity it will be considered the greybody spectrum (\ref{gbspect}) without the phase space factor, which is equal for each case, therefore without loss of generality one can redefine the final observable as 
\be\label{Gfactor}
\gamma^{\text{grey}}_{N}(E_{k}/T_{eff})= C_{N}(\{g_{n}\}.\{g'_{n'}\}) {\left({E_{k}\over T_{eff}}\right)^{r_{N}(\{g_{n}\},\{g'_{n'}\})}\over e^{\nu_{N}(\{g_{n}\},\{g'_{n'}\}){E_{k}\over T_{eff}}}-1}
\ee

\section{Results of random generated spectra}\label{se3}
Let's start by considering a simple example in which only a single microstate can decay, in particular let's consider the specific microstate ${\small \Pi^{(100)}_{\pi_{1}(g_{n})}}$ of the level $N=100$ given by
\be\label{fbk22}
\pi_{1}(g_{n})=\{g_{1}=1,g_{2}=1,g_{5}=2,g_{7}=1,g_{8}=1,g_{9}=1,g_{10}=1,g_{25}=1,g_{28}=1\}
\ee 
The first energy point of the spectrum is given by the decay of $\Pi^{(100)}_{\pi_{1}(g_{n})}$ into a generic microstate of the level $N'=N-1$ and corresponds to $E_{k}=0$. The fact that $E_{k}=0$ is constrained by the kinematics of the emitted tachyonic particle which make the first point of the spectrum to be trivial. 

\begin{figure}[h!]
\centering
\includegraphics[scale=0.50]{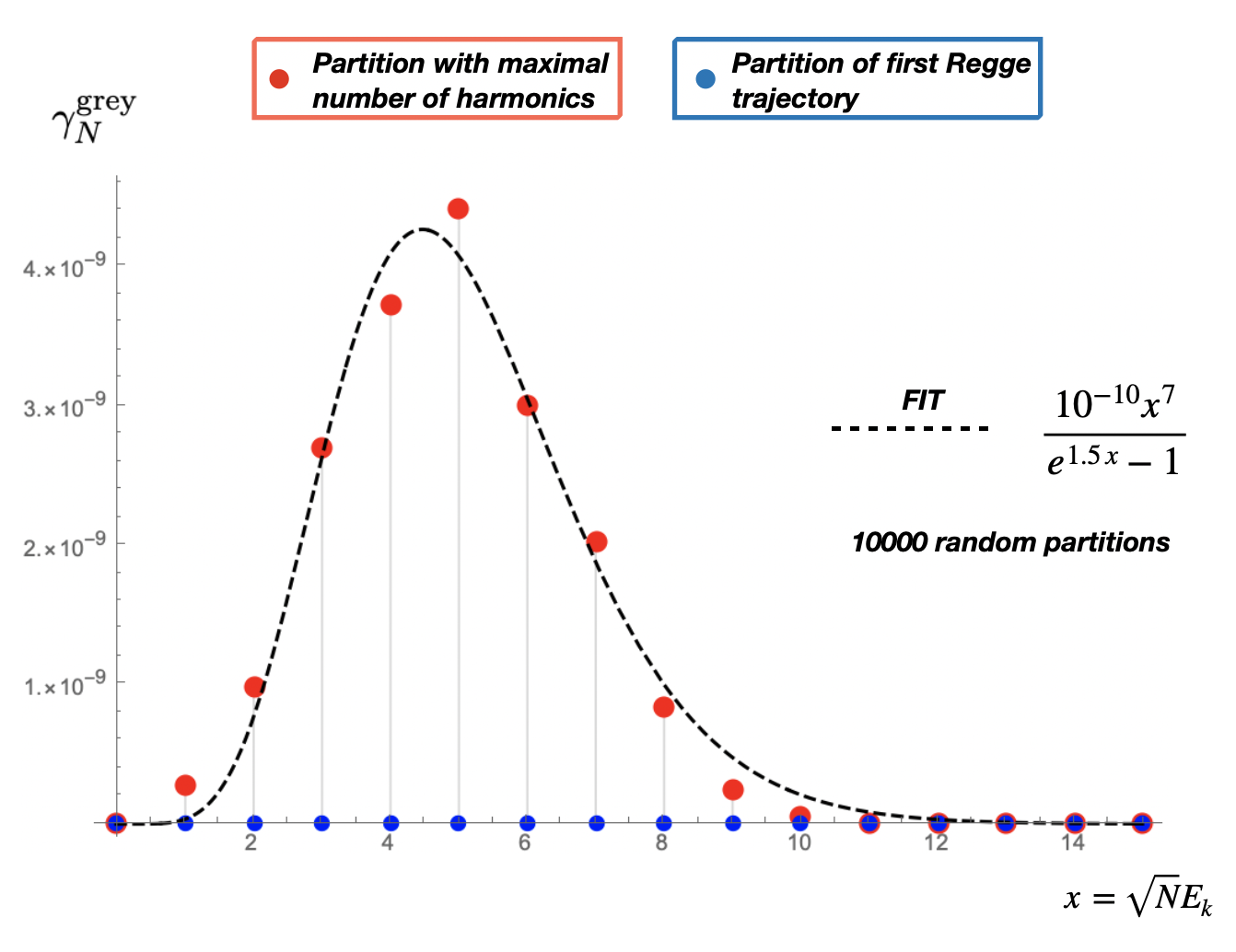}
\caption{Spectra of microstates of $N=100$. Comparison between the spectrum of the first Regge trajectory (blue points) and the spectrum of the state with the maximal number of harmonics (red points). The black dashed line represents the fitted behavior of the red spectrum.}\label{thermFRt}
\end{figure}

The first non trivial energy point of the spectrum is given by the decay of $\Pi^{(100)}_{\pi_{1}(g_{n})}$ into a generic microstate of the level $N'{=}N{-}2$, and the consecutive points are obtained from $N'{=}N{-}3$,  $N'{=}N{-}4$ and so on. Since the HES of the level $N'$ 
is generically composed by many degenerate microstates, one can introduce the intrinsic random nature of the process taking the average over different sets of random microstates for each energy point of the spectrum. 
\begin{figure}[h!]
\centering
\includegraphics[scale=0.55]{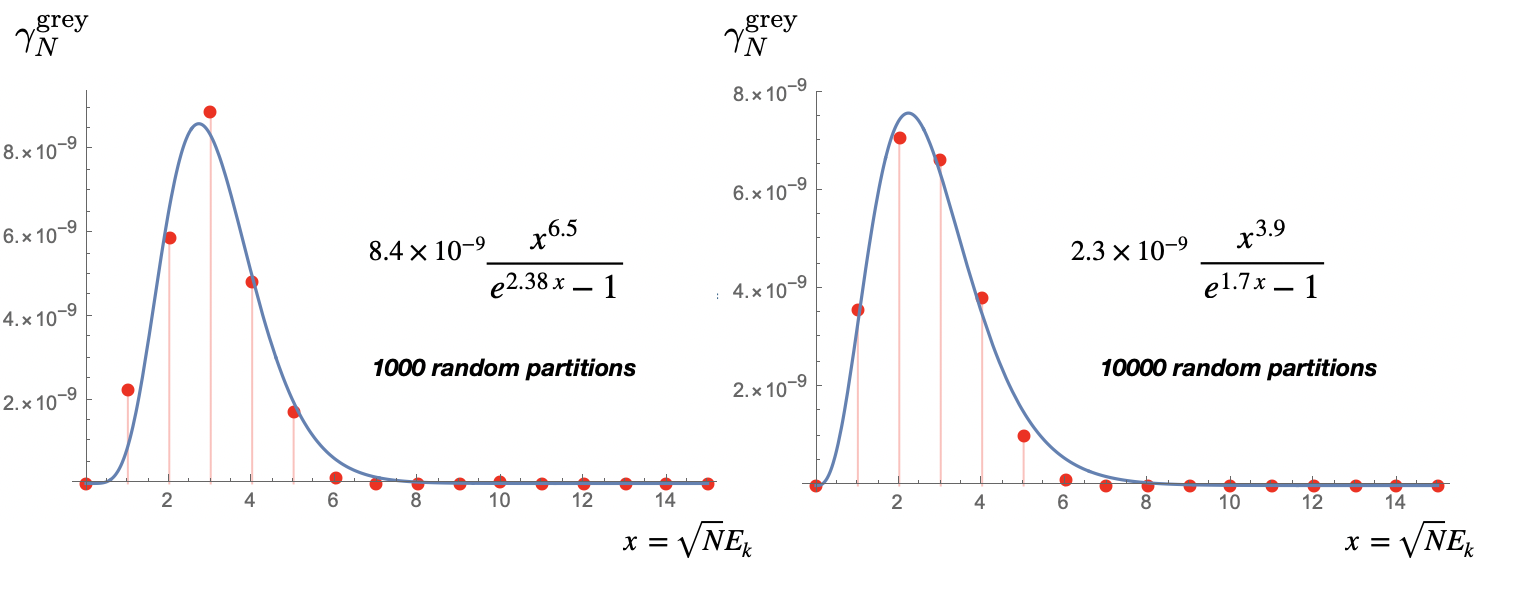}
\caption{Thermal spectrum of the representative microstate $\Pi^{(100)}_{\pi_{1}(g_{n})}$ computed for two different random sample of microstates of $N'$.}\label{SpecN100}
\end{figure}
In order to make clear the relation between chaos and thermalization one can adopt the same logic of (fig.\ref{figN100}), in particular one can compare the spectrum of the microstate with the maximal number of harmonics and the extreme case of the first Regge trajectory (fig.\ref{thermFRt}). 
As expected the spectrum of a state of the FRtj is not thermal.

\begin{figure}[h!]
\centering
\includegraphics[scale=0.4]{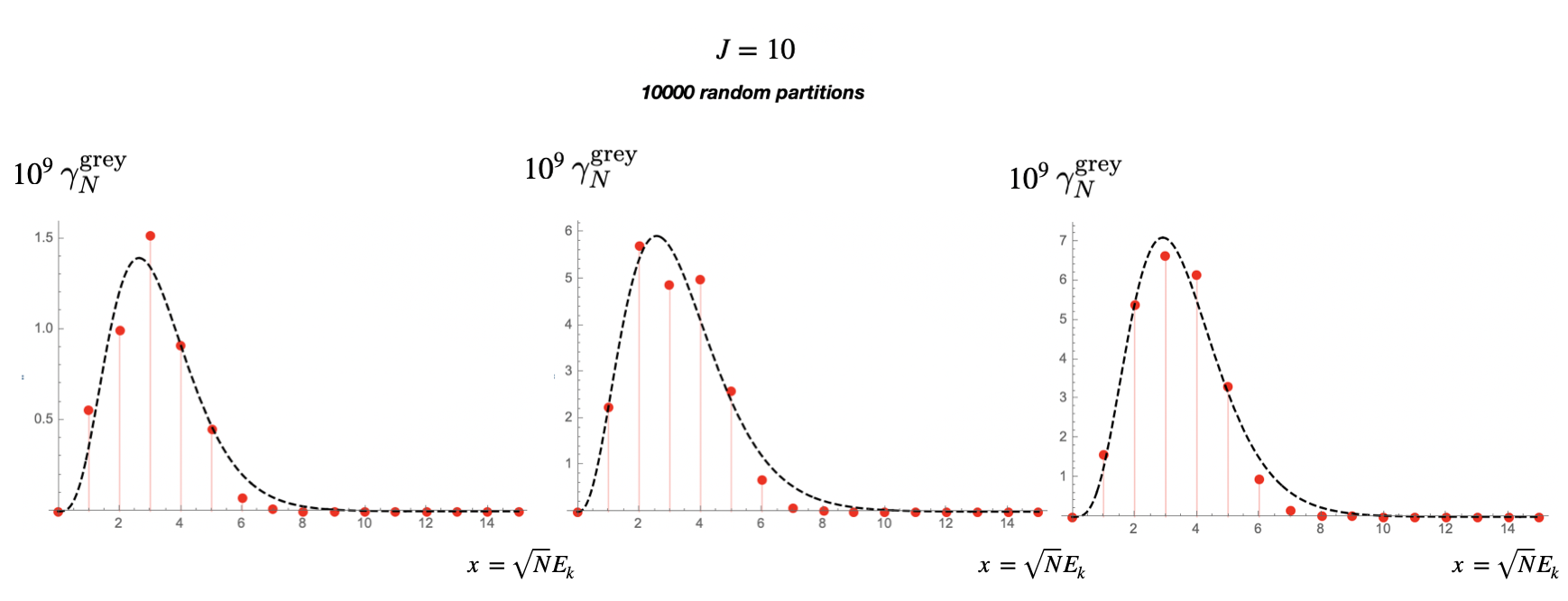} \includegraphics[scale=0.4]{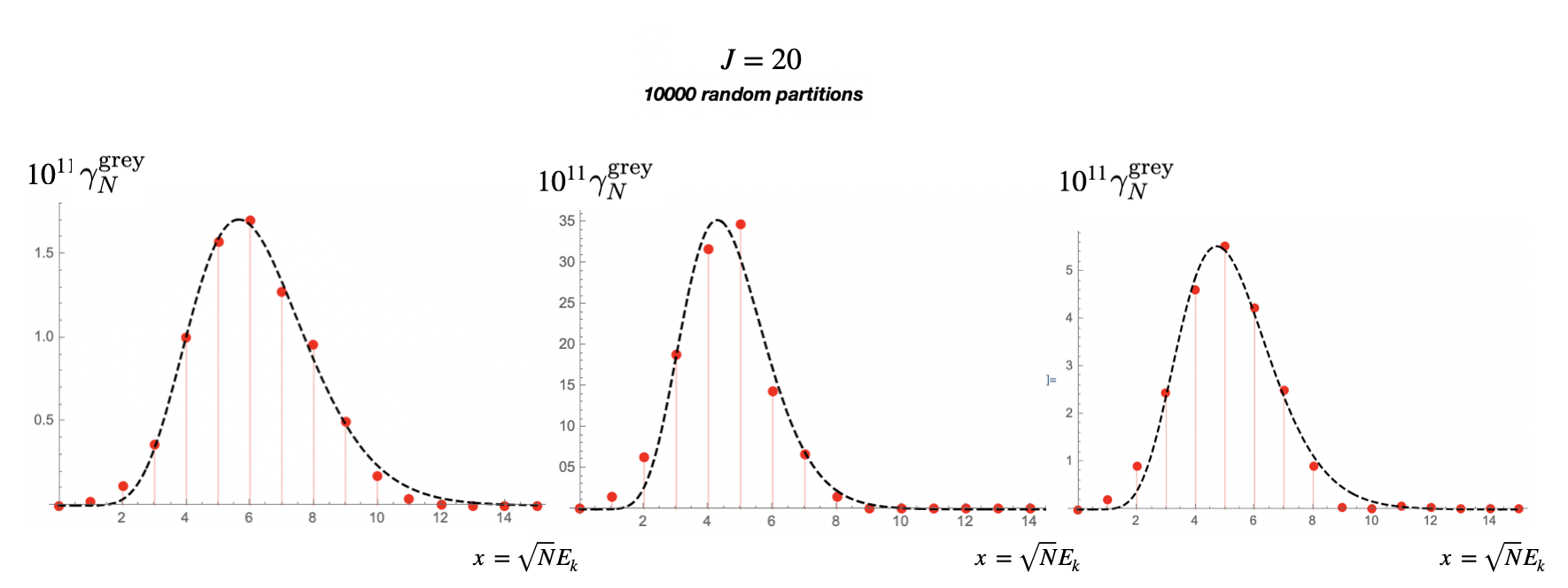} \includegraphics[scale=0.4]{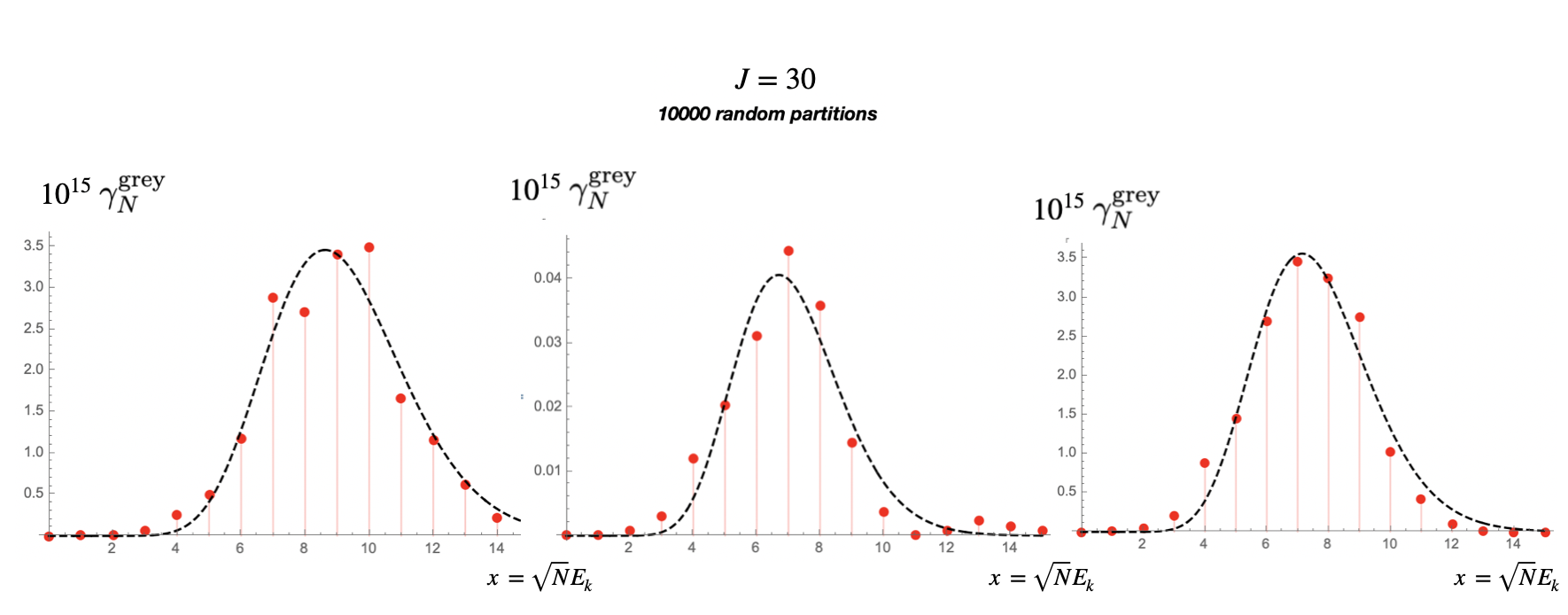} \includegraphics[scale=0.4]{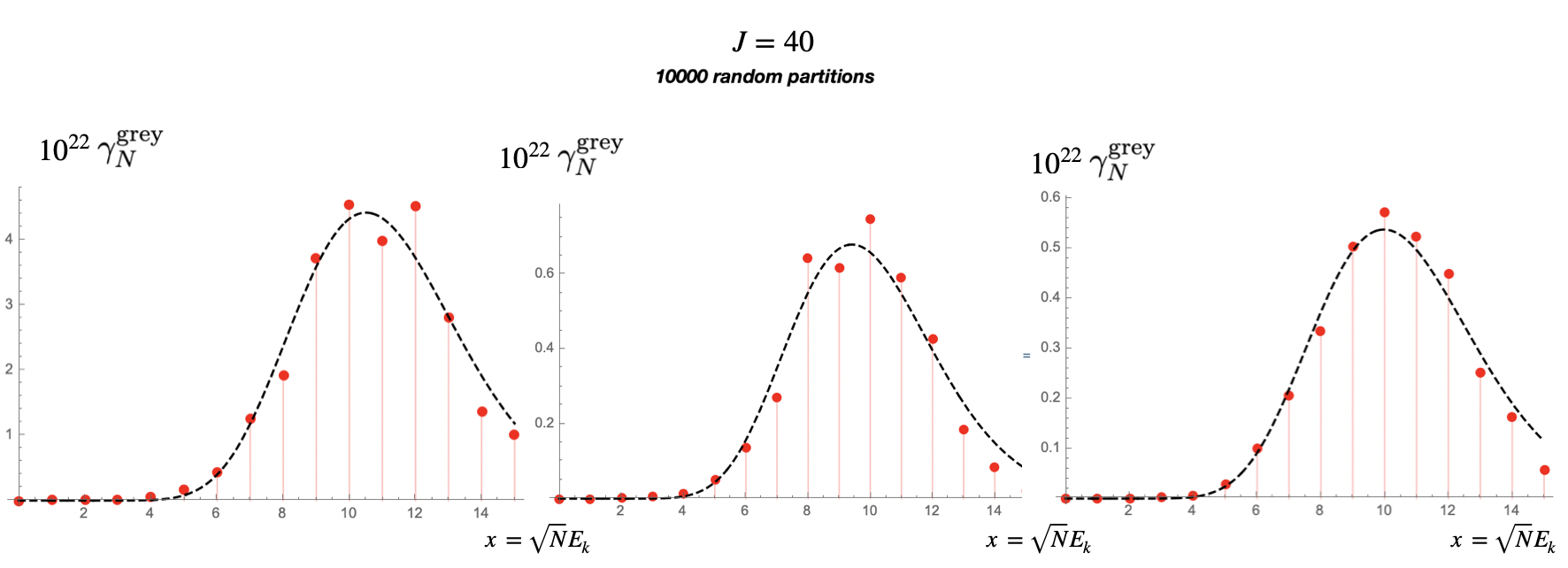}
\caption{Some examples of thermal spectra as a function of the occupation number $J$.}\label{Jtherm}
\end{figure}

A complementary picture of the thermal spectrum of string microstates is given in fig.\ref{Jtherm}, where many spectra are presented, classified by different microstates and values of the occupation number $J$. The specific parameters of (\ref{Gfactor})  are reported in Tab.\ref{table:N=10}.

\begin{table}[h!]
\centering
\begin{tabular}{c | c || c| c| c|}
\hline\hline
$J$ &Microstate & $C_{N}$ & $r_{N}$ & $\nu_{N}$ \\ [0.5ex] 
\hline
& $\{g_{2}{=}3,\,g_{12}{=}2,\,g_{3}{=}g_{10}{=}g_{13}{=}g_{20}{=}g_{24}{=}1\}$ &$1.49\times10^{-9}$&4.38&1.64\\
10 & $\{g_{2}{=}g_{3}{=}g_{4}{=}g_{5}{=}g_{6}{=}g_{13}{=}g_{15}{=}g_{16}{=}g_{17}{=}g_{19}{=}1\}$ &$4.90\times 10^{-9}$&4.70&1.60\\
& $\{g_{13}{=}2,\,g_{1}{=}g_{2}{=}g_{4}{=}g_{6}{=}g_{7}{=}g_{12}{=}g_{19}{=}g_{23}{=}1\}$ &$6.23\times 10^{-9}$&3.53&1.33\\
\hline
&$\{g_{2}{=}6,\,g_{3}{=}3,\,g_{5}{=}g_{9}{=}2,\,g_{1}{=}g_{4}{=}g_{6}{=}g_{7}{=}g_{8}{=}g_{12}{=}g_{13}{=}1\}$ &$1.27\times 10^{-14}$&9.84&1.74 \\
20 &$\{g_{2}{=}8,\,g_{7}{=}2,\,g_{1}{=}4,\,g_{3}{=}g_{4}{=}g_{11}{=}g_{12}{=}g_{16}{=}g_{20}{=}1\}$ &$2.08\times 10^{-14}$&11.30&2.64\\
&$\{g_{1}{=}g_{2}{=}3,\,g_{3}{=}4,\,g_{4}{=}3,\,g_{8}{=}2,\,g_{5}{=}g_{6}{=}g_{7}{=}g_{12}{=}g_{21}\}$ &$28.9\times 10^{-14}$&9.52&2.01\\
\hline
&$\{g_{3}{=}g_{4}{=}g_{6}{=}3,\,g_{1}{=}8,\,g_{2}{=}6,\,g_{5}{=}5,g_{7}{=}g_{9}{=}1\}$ &$1.00\times 10^{-23}$&17.02&1.97 \\
30 &$\{g_{1}{=}8,\,g_{2}{=}11,\,g_{3}{=}g_{6}{=}3,\,g_{8}{=}2,\,g_{4}{=}g_{5}{=}g_{18}{=}1\}$ &$7.51\times 10^{-24}$&17.18&2.56\\
&$\{g_{1}{=}11,\,g_{2}{=}7,\,g_{3}{=}3,\,g_{4}{=}g_{7}{=}2,\,g_{5}{=}g_{6}{=}g_{8}{=}g_{11}{=}g_{14}{=}1\}$ &$9.74\times 10^{-22}$&15.63&2.18\\
\hline
&$\{g_{1}{=}15,\,g_{2}{=}9,\,g_{3}{=}7,\,g_{4}{=}4,g_{5}{=}3,\,g_{6}{=}g_{9}{=}1\}$ &$7.44\times 10^{-33}$&18.35&1.74 \\
40 &$\{g_{1}{=}7,\,g_{2}{=}10,\,g_{3}{=}g_{4}{=}4,\,g_{6}{=}3,\,g_{5}{=}g_{12}{=}1\}$ &$6.22\times10^{-32}$&16.75&1.78\\
&$\{g_{1}{=}17,\,g_{2}{=}9,\,g_{3}{=}7,\,g_{6}{=}3,g_{7}{=}2,\,g_{4}{=}g_{8}{=}1\}$ &$6.27\times10^{-32}$&16.75&1.78\\
\hline
\end{tabular}
\caption{Table of specific microstates and parameters of (\ref{Gfactor}) relative to the spectra of fig.\ref{Jtherm}. }
\label{table:N=10}
\end{table}

\section{Conclusion and future directions}
In section \ref{se1} we have studied the interplay between classical string configurations and quantum string configurations, where the latter are essentially the microstates of the mass degeneracy of the HES. We have observed how the chaotic nature of HES interactions has a common pattern with the shape of classical string profiles, which is provided by the number of harmonics that characterizes the microstate. We have presented explicit results for a representative HES at mass level $N=100$, but the same systematics holds for different mass levels. From the point of view of the analysis of the chaotic information present in HES interactions, one can quantitatively improve the measure of chaos with additional parametrization of the information content based on modern techniques of quantum information theory \cite{Witten:2018zva}-\cite{Erbin:2022rgx}

 In section \ref{se2} we have analyzed how the thermal nature of the decay amplitude is intimately related to its chaotic nature, in particular we have observed that the chaotic analytical structure, which appear as a non trivial dressing factor, originates a Boltzmann factor which encodes the thermal information of the decay process. Exploiting the exact analytical result of the decay process of a generic HES that decays into a less excited, yet generic HES, through the emission of scalar particle, such as the tachyon, we have computed the energy spectrum of HES. Starting from a definite microstate of the level $N$, we computed the average over random microstates of the level $N'$, for many different values of $N'$. In particular the thermal spectra we found are originated by the randomization of all the possible final microstates. This prescription results connected with the random walk nature of HES through the emergence of an effective temperature which scales as $N^{-{1\over 2}}$, which is the inverse of the characteristic size of HES. Such temperature is the result of the numerical analysis performed in section \ref{se3}. 
 
To be more precise, in section \ref{se3} we have presented numerical results of the extracted spectra for many different microstates of the representative level $N=100$, but the same holds for higher levels. Lower levels, $N<100$, have less accessible energy points of interest.
 We numerically confirmed that for the expression (\ref{Gfactor}) the coefficient $\nu_{N}\simeq O(N^{0})$ while $C_{N}$ and $r_{N}$ are sensitive to the microstate structure (as we have seen from table \ref{table:N=10}). As a result we observed how a string of the first Regge trajectory deviates from the thermal spectrum, in fact it is not enough excited to produce a chaotic interaction leading to a thermal behavior, while a generic HES produces such behavior (fig.\ref{thermFRt}).  We also observed a non trivial dependence of the spectra on the occupation number $J$ that we want to quantitatively address in future works, together with a fully quantitative computation of the spectrum of degenerate HES, which means the computation of the linear combination of the same spectra we computed, but for all the possible microstate of the level $N$.

 The chaotic nature of the decay process together with the non trivial dependence of final microstates gave rise to a greybody emission spectrum with an effective temperature $T_{eff}=T_{H}/\sqrt{N}$, which is different from the temperature of the string/BH transition \cite{Horowitz:1996nw}. In fact we recovered the characteristic behavior of the Hawking temperature which is expected to be proportional to the inverse of the mass of the decaying state. A possible interpretation of such result can be connected to an enhancement of the effective string Schwarzschild radius, due to the random walk nature of HES interactions modeled by the explicit introduction of the microstates dependence. The chaotic nature of HES interactions \cite{Gross:2021gsj}-\cite{ChaosScatt}, and also the associated thermal nature suggest a non trivial spatial distribution of HES, which can be probed in a scattering experiment similar to the analysis in \cite{Mitchell:1990cu}. It is well known that an HES can be described as random walk of interactions \cite{Kruczenski:2005pj}-\cite{Damour:1999aw}, and than one can measure the precise microstates structure of the spatial distribution of HES studying the effective horizon, for example, probed in HES Compton-like scattering processes. Implementing random surface techniques \cite{Tan:1980zh}-\cite{Charles:2018oob} to the HES form factors one can obtain a complementary picture of the HES nature where chaotic and thermal effects can be matched with the effective HES horizon governed by the superposition of microstates. We leave this investigation for future works.
 
Alternatively quite recently it was proposed a technique to resolve the spatial distribution of strings \cite{Hashimoto:2022ugt} and also a connected chaotic analysis of the HES Compton scattering \cite{Hashimoto:2022bll}, based on the principle of transient chaos.

The understanding of the intrinsic structure of HES along with a complete picture of HES interactions can be very useful in studying deep microscopical connections between thermalization and chaos. The statistical non trivial nature of HES provides a very rich physical system which still deserves to be further explored.

\section*{Acknowledgements}
I would like to thank M.~Bianchi, G.~Rossi,  J.~Sonnenschein, D.~Weissman, V.~Rosenhaus, D.~Gross, B.~Sundborg, A.~Tseytlin, G.~Di~Russo,  A.~Guerrieri and V.~Niarchos for valuable discussions and comments.
I would like also to thank The Graduate Center, CUNY for the hospitality during the completion of the manuscript.

\vspace{3cm}
\begin{appendix}\label{AApp}
\section{Highly excited string decay: $H_{N}\Rightarrow H_{N'}+T$} \label{appA}
The present appendix concerns a detailed review and new insights about the decay process computation based on \cite{Firrotta:2022cku}. In particular we have presented the analytical setup with the main steps of the computation and also it will be discussed the behavior of the decay rate in the thermalization region, which is reached when the ratio between the energy of the emitted state and the mass of the decaying state is enough small in such a way that the energy loss of the decaying state is smooth.
\begin{figure}[h!]
\centering
\includegraphics[scale=0.55]{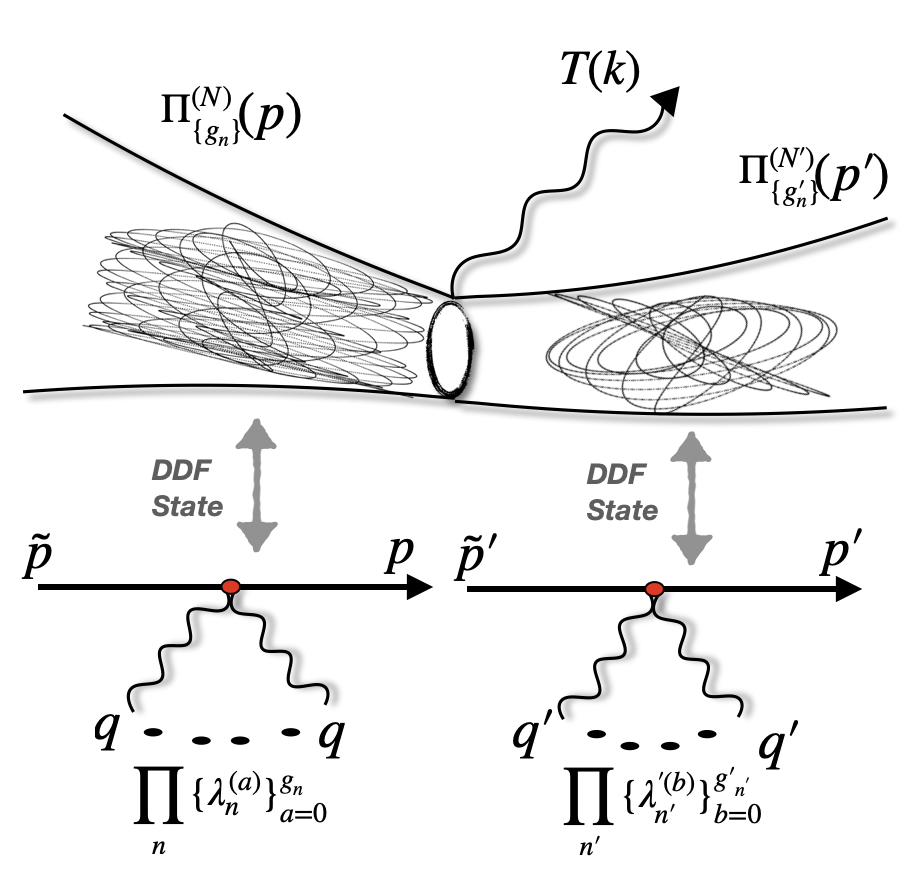}
\caption{Picture of the decay amplitude where a representative microstate $\Pi_{\{g_{n}\}}^{(N)}$ decays into $\Pi_{\{g'_{n'}\}}^{(N')}$ emitting a tachyon $T$. The DDF structure of the microstates is also depicted below the decay process.} 
\label{figHHTDDF}
\end{figure}
Following the picture in (fig.\ref{figHHTDDF}) the kinematics of the process is parametrized as follows
\be\label{fbk23}
p=\sqrt{2N-2}(1,\vec{0})\,,\quad p'=-(E',\omega\sin\theta,\omega\cos\theta,\vec{0})\,,\quad k=(-E_{k},\omega\sin\theta,\omega\cos\theta,\vec{0})
\ee
\be\label{fbk24}
q=-{(1,0,1,\vec{0})\over \sqrt{2N-2}}\,,\quad \lambda={(0,1,0,\vec{\Lambda})\over \sqrt{1+|\vec{\Lambda}|^{2}}}\,;\quad q'=-{(1,0,1,\vec{0})\over \omega\cos\theta-E'}\,,\quad \lambda'={(0,1,0,\vec{\Lambda}')\over \sqrt{1+|\vec{\Lambda}'|^{2}}}
\ee
where the momenta $\tilde{p}$ and $\tilde{p}'$ of (fig.\ref{figHHTDDF}) are tachyonic DDF reference momenta, while $p=\tilde{p}-\sum_{n}ng_{n}\, q$ and $p'=\tilde{p}'-\sum_{n'}n'g'_{n'}\, q'$ are the total momenta of the microstates $\Pi_{\{g_{n}\}}^{(N)}$ and $\Pi_{\{g'_{n'}\}}^{(N')}$. The nice feature of the DDF formalism relies in the direct identification of the final BRST vertex operator corresponding to the microstate, where its structure is modeled by the number of DDF photon insertions with polarizations $\lambda_{n}$, $\lambda'_{n}$ and momenta $q$ and $q'$. The photon insertions are in exact correspondence with each harmonic and every excitation number, exactly reproducing the action of creation operators.

\subsection{Decay amplitude}
Let's start by introducing the generating amplitude of all the possible decay processes of $H_{N}\Rightarrow H_{N'}+T$ discriminated by all the possible microstate $\Pi^{(N)}_{\{g_{n}\}}$ and $\Pi^{(N')}_{\{g_{n'}\}}$ originated from the partition of integers $N=\sum_{n}ng_{n}$ and $N'=\sum_{n'}n'g_{n'}$ 
\be\label{fbk25}
\begin{split}
{\cal A}_{gen}=\exp\Bigg(&\sum_{n;\,a_{n}=1}^{g_{n}}  J_{n}^{(a_{n})}{\cdot}V_{n} +{J}_{n}^{'(a_{n})}{\cdot}V'_{n} + \\
&\sum_{n,m ; \,a_{n},b_{m}}^{g_{n},g_{m}} J^{(a_{n})}_{n}{\cdot}J^{(b_{m})}_{m} W_{m,n}+J'^{(a_{n})}_{n}{\cdot}J'^{(b_{m})}_{m} W'_{m,n} +J^{(a_{n})}_{n}{\cdot}J'^{(b_{m})}_{m} M_{m,n}  \Bigg)
\end{split}
\ee
where all the interaction terms are classified as follows
\be \label{V1}
{\cal V}_{n}^{\mu}={p'}^{\mu}V_{n}={p'}^{\mu}{(-)^{n+1}\over \Gamma(n) }{(1+nq{\cdot}p')_{n{-}1}}  \,, \quad {{\cal V}'}^{\mu}=p^{\mu}V'_{n}={p^{\mu}\over \Gamma(n)} (1+nq'{\cdot}p)_{n{-}1} 
\ee
\be\label{V2}
W_{n,m}={n\,m\over n+m}  {(1+q{\cdot}p')\,q{\cdot}p'}\, V_{n} \,V_{m} \, , \quad  W'_{n,m}={n\,m\over n+m}{(1+q'{\cdot}p)\, q'{\cdot}p}\, V'_{n} \, V'_{m}  
\ee
\be\label{M12}
M_{n,m}=-{n m (1+q{\cdot}p')\over m + n q{\cdot}p'} V_{n} V'_{m}
\ee
Any particular decay amplitude can be obtained by operating with derivative combinations representing the projection on the single amplitude with the desired states identified by the partition set $\{g_{n}\}$ and $\{g_{n'}\}$ as 
\be\label{fbk26}
{\cal A}_{\Pi^{N}_{\{g_{n}\}},\Pi^{N'}_{\{g_{n'}\}}}=\prod_{n}\prod_{a_{n}=1}^{g_{n}} \zeta_{n}^{(a_{n})}{\cdot}{d\over dJ_{n}^{(a_{n})}}\,\prod_{n'}\prod_{a_{n'}=1}^{g_{n'}} {\zeta'}_{n'}^{(a_{n'})}{\cdot}{d\over d{J'}_{n'}^{(a_{n'})}}\,{\cal A}_{gen}\Bigg|_{J=J'=0}
\ee
where all the polarizations $\zeta_{n\,,\mu}^{(a_{n})}=\lambda^{(a_{n})}_{n\,,\mu}-\lambda^{(a_{n})}_{n}{\cdot}pq_{\mu}$ and ${\zeta'}_{n'\,,\mu}^{(a_{n'})}={\lambda'}_{n'\,,\mu}^{(a_{n'})}-{\lambda'}_{n'}^{(a_{n'})}{\cdot}p'q'_{\mu}$ are independent.

By considering a kinematical setup where the decaying string is at rest, one can compute the relevant scalar product $q{\cdot}p'$ that encodes the partition dependence of the interacting states. In particular 
\be\label{fbk27}
E'=M_{N}-E_{k}\,,\quad \omega= \sqrt{E_{k}^{2}-M_{T}^{2}}\,,\quad E_{k}={N-N'-1\over \sqrt{2N-2}}
\ee
and using the kinematics one finds
\be\label{fbk28}
q{\cdot}p'={1\over q'{\cdot}p}=-{E'-\omega \cos\theta\over M_{N}}= -1+{E_{k}\over M_{N}}+{\sqrt{2}\over M_{N}}\sqrt{1+{E_{k}^{2}\over 2}}\cos\theta
\ee
Without loss of generality one can chose $\theta=\pi/2$, in fact the final observable will be the decay rate, so when the modulus square of the amplitude is considered, the exact spherical symmetry is restored. It means that one can freely fix the value of $\theta$ without loss of generality. With this choice one has
\be\label{fbk29}
q{\cdot}p'= -1+{E_{k}\over M_{N}}
\ee 
In this framework one can easily analyze the non trivial contributions in (\ref{V1}), (\ref{V2}) and (\ref{M12}):
\be\label{fbk30}
V_{n}=(-)^{n+1}{(1+nq{\cdot}p')_{n{-}1}\over \Gamma(n)}= {(-)^{n+1}\Gamma\left(n {E_{k}\over M_{N}}  \right)\over \Gamma(n)\Gamma\left(1-n(1-{E_{k}\over M_{N}}) \right)}
\ee
this is the oscillating function that generates chaos, in fact it can be written as
\be\label{fbk31}
V_{n}={1\over \pi\,\Gamma(n)}{\Gamma\left(n {E_{k}\over M_{N}}  \right) \Gamma\left(n-n{E_{k}\over M_{N}} \right)}\sin\left(n\pi{E_{k}\over M_{N}} \right)
\ee
The other term to analyze is
\be\label{fbk32}
W_{n,m}={n\,m\over n+m}  {(1+q{\cdot}p')\,q{\cdot}p'}\, V_{n} \,V_{m}={n\,m\over n+m}  {E_{k}\over M_{N}} \left( {E_{k}\over M_{N}} -1\right)\, V_{n} \,V_{m}
\ee
and also there is 
\be\label{fbk33}
V'_{n}={(1+nq'{\cdot}p)_{n{-}1}\over \Gamma(n)}={\Gamma\left(-n {E_{k}\over M_{N}}  \right)\over \Gamma(n)\Gamma\left(1-n(1+{E_{k}\over M_{N}}) \right)}
\ee
where both numerator and denominator are oscillating
\be\label{fbk34}
V'_{n}={\Gamma\left(n+n{E_{k}\over M_{N}} \right)\over \Gamma(n) \Gamma\left(1+n {E_{k}\over M_{N}}  \right) } {\sin\left( n\pi+n\pi{E_{k}\over M_{N}} \right)\over \sin\left( n\pi{E_{k}\over M_{N}} \right)}=(-)^{n}{\Gamma\left(n+n{E_{k}\over M_{N}} \right)\over \Gamma(n) \Gamma\left(1+n {E_{k}\over M_{N}}  \right) }
\ee
then there is 
\be\label{fbk35}
W'_{n,m}={n\,m\over n+m}  {(1+q'{\cdot}p)\,q'{\cdot}p}\, V'_{n} \,V'_{m}={n\,m\over n+m}  {E_{k}\over M_{N}} \left( {E_{k}\over M_{N}} +1\right)\, V'_{n} \,V'_{m}
\ee 
finally the mixed term
\be\label{fbk36}
M_{n,m}=-{n m (1+q{\cdot}p')\over m + n q{\cdot}p'} V_{n} V'_{m}=-{n m {E_{k}\over M_{N}}\over m - n+n{E_{k}\over M_{N}}} V_{n} V'_{m}
\ee
To sum up one has the following structures
\be\label{fbk37}
{\cal V}_{n}^{\mu}={p'}^{\mu}V_{n}(E_{k})\,, \quad {{\cal V}'}_{n}^{\mu}=p^{\mu}V'_{n}(E_{k})\,,\quad M_{n,m}=\mu_{n,m}(E_{k})V_{n}(E_{k})V'_{m}(E_{k})
\ee
\be\label{fbk38}
W_{n,m}=w_{n,m}(E_{k})V_{n}(E_{k})V_{m}(E_{k})\,,\quad  W'_{n,m}=w'_{n,m}(E_{k})V'_{n}(E_{k})V'_{m}(E_{k})
\ee
where
\be\label{fbk39}
\mu_{n,m}(E_{k})=-{n m {E_{k}\over M_{N}}\over m - n+n{E_{k}\over M_{N}}}\,,\quad w_{n,m}(E_{k})={n\,m\over n+m}  {E_{k}\over M_{N}} \left( {E_{k}\over M_{N}} -1\right)
\ee
\be\label{fbk40}
w'_{n,m}(E_{k})={n\,m\over n+m}  {E_{k}\over M_{N}} \left( {E_{k}\over M_{N}} +1\right)
\ee
Finally the general decay amplitude can be written as
\be\label{fbk41}
{\cal A}_{\Pi^{N}_{\{g_{n}\}},\Pi^{N'}_{\{g_{n'}\}}}= {\cal P}_{\Pi_{N},\Pi_{N'}}\left(\zeta_{n}^{(a)}\,,\zeta_{n'}^{\,_{'}(a')}\,,p\,,p'\,, w_{n,m}\,,w'_{n',m'}\,,\mu_{n,n'}\right)\, \prod_{n}\big(V_{n}\big)^{g_{n}} \prod_{n'}\big({V_{n'}'}\big)^{g_{n'}}
\ee
where ${\cal P}_{\Pi_{N},\Pi_{N'}}$ is a polynomial that depends on the partition of $N$ and $N'$ and it can be obtained by the generating function
\be \label{fbk42}
\begin{split}
{\cal P}_{gen}=\exp\Bigg(&\sum_{n;\,a_{n}=1}^{g_{n}}  J_{n}^{(a_{n})}{\cdot}p' +\sum_{n;\,a_{n}=1}^{g_{n}} {J}_{n}^{'(a_{n})}{\cdot}p+\sum_{n,m ; \,a_{n},b_{m}}^{g_{n},g_{m}} J^{(a_{n})}_{n}{\cdot}J'^{(b_{m})}_{m} \mu_{m,n} \\
&\sum_{n,m ; \,a_{n},b_{m}}^{g_{n},g_{m}} J^{(a_{n})}_{n}{\cdot}J^{(b_{m})}_{m} w_{m,n}+\sum_{n,m ; \,a_{n},b_{m}}^{g_{n},g_{m}} J'^{(a_{n})}_{n}{\cdot}J'^{(b_{m})}_{m} w'_{m,n}   \Bigg)
\end{split}
\ee
the important thing to note is that the oscillating terms $V_{n}$ and $V'_{n}$  are factorized from the polynomial. They are the same factors that produce chaos and that contain most of the information about the microstate dependence.

\subsection{Decay rate}
The general structure of the absolute value square of the amplitude is given by
\be\label{fbk43}
\Big|{\cal A}_{\Pi^{N}_{\{g_{n}\}},\Pi^{N'}_{\{g_{n'}\}}}\Big|^{2}=\Big|{\cal P}_{\Pi_{N},\Pi_{N'}}\left(\zeta_{n}\,,\zeta'_{n'}\,,p\,,p'\,, w_{n,m}\,,w'_{n',m'}\,,\mu_{n,n'}\right)\Big|^{2}\, \prod_{n}\big(V^{2}_{n}\big)^{g_{n}} \prod_{n'}\big({{V_{n'}'}^{2}}\big)^{g_{n'}}
\ee
where the polynomial is generated by
\be\label{fbk44}
{\cal P}_{\Pi_{N},\Pi_{N'}}=\prod_{n}\prod_{a_{n}=1}^{g_{n}} \zeta_{n}^{(a_{n})}{\cdot}{d\over dJ_{n}^{(a_{n})}}\,\prod_{n'}\prod_{a_{n'}=1}^{g_{n'}} {\zeta'}_{n'}^{(a_{n'})}{\cdot}{d\over d{J'}_{n'}^{(a_{n'})}}\,{\cal P}_{gen}\Bigg|_{J=J'=0}
\ee
Taking the absolute value square and using the completeness relations
\be\label{fbk45}
\sum_{pol}\zeta_{n}^{(a)\mu}\zeta_{n}^{*(a)\mu}={\cal L}^{(a)\mu\nu}\,, \quad \sum_{pol}{\zeta'}_{n}^{(a)\mu}{\zeta'}_{n}^{*(a)\mu}={{\cal L}'}^{(a)\mu\nu}
\ee
where the superscript $(a)$ refers to different independent polarizations, with the explicit completeness structures
\be\label{fbk46}
{\cal L}^{(a)\mu\nu}=\eta^{\mu\nu}-2p^{(\mu}q^{\mu)}+p^{2}q^{\mu}q^{\nu}\,, \quad {{\cal L}'}^{(a)\mu\nu}=\eta^{\mu\nu}-2p'^{(\mu}q'^{\mu)}+p'^{2}q'^{\mu}q'^{\nu}
\ee
one can study the relevant contributions of the polynomial, in particular the first one yields 
\be\label{fbk47}
p'{\cdot}{\cal L}{\cdot}p'={p'}^{2}-2p{\cdot}p'\,q{\cdot}p'+p^{2}(q{\cdot}p')^{2} \,, \quad \text{with}\,\,\, q{\cdot}p'=-1+{E_{k}\over M_{N}}
\ee
that can be written as
\be\label{fbk48}
p'{\cdot}{\cal L}{\cdot}p'=(p+p')^{2}-2p{\cdot}p' {E_{k}\over M_{N}} - 2p^{2}{E_{k}\over M_{N}}+ O(E_{k}^{2})
\ee
using $p{\cdot}p'=M_{N}E'=M_{N}^{2}-M_{N}E_{k}$, $p^{2}=-M_{N}^{2}$ and $(p+p')^{2}=k^{2}$ one has
\be\label{fbk49}
p'{\cdot}{\cal L}{\cdot}p'=k^{2}-2M_{N}E_{k}+2E_{k}^{2}+2M_{N}E_{k}=\omega^{2}+O(E_{k}^{2})\simeq -M_{T}^{2}=2
\ee
The second contribution is given by
\be\label{fbk50}
p{\cdot}{\cal L}'{\cdot}p=p^{2}-2p{\cdot}p' q'{\cdot}p+{p'}^{2}(q'{\cdot}p)^{2}\,, \quad \text{with}\,\,\,q'{\cdot}p=-1-{E_{k}\over M_{N}}
\ee
which yields
\be\label{fbk51}
p{\cdot}{\cal L}'{\cdot}p=(p+p')^{2}+2p{\cdot}p'{E_{k}\over M_{N}}+2{p'}^{2}{E_{k}\over M_{N}} 
\ee
still using $p{\cdot}p'=M_{N}E'=M_{N}^{2}-M_{N}E_{k}$, ${p'}^{2}=-{M}_{N'}^{2}$ and $(p+p')^{2}=k^{2}$ one has
\be\label{fbk52}
p{\cdot}{\cal L}'{\cdot}p=k^{2}+2M_{N}E_{k}-2E_{k}^{2}-{M_{N'}^{2}\over M_{N}^{2}} 2M_{N}E_{k} 
\ee
and finally noting that
\be\label{fbk53}
{M_{N'}^{2}\over M_{N}^{2}}=1-{E_{k}\over M_{N}} + O(1/M_{N}^{2})
\ee
the expression becomes
\be\label{fbk54}
p{\cdot}{\cal L}'{\cdot}p=k^{2}=2
\ee
Using the same steps the last term can be written as
\be\label{fbk55}
p'{\cdot}{\cal L}{\cdot}{\cal L}'{\cdot}p=M_{N}E' + M_{N}^{2}q{\cdot}p'+M_{N'}^{2}q'{\cdot}p=2(1+E_{k}/M_{N}) + O(E_{k}^{2})
\ee

This is a map from ${\cal P}_{\Pi_{N},\Pi_{N'}}$ and its absolute value square, where the latter can be recasted in a new polynomial ${\cal X}_{\Pi_{N},\Pi_{N'}}$ 
\be\label{fbk56}
\Big|{\cal P}_{\Pi_{N},\Pi_{N'}}\Big|^{2}={\cal X}_{\Pi_{N},\Pi_{N'}}\left(p'{\cdot}{\cal L}^{(a)}{\cdot}p',\,p{\cdot}{{\cal L}'}^{(a)}{\cdot}p,\,p'{\cdot}{\cal L}^{(a)}{\cdot}{{\cal L}'}^{(b)}{\cdot}p,\, w_{n,m},\,w'_{n',m'},\,\mu_{n,n'} \right)
\ee
Now let's compare the amplitude with its mod square
\be\label{fbk57}
{\cal A}_{\Pi^{N}_{\{g_{n}\}},\Pi^{N'}_{\{g_{n'}\}}}= {\cal P}_{\Pi_{N},\Pi_{N'}}\, \prod_{n}\big(V_{n}\big)^{g_{n}} \prod_{n'}\big({V_{n'}'}\big)^{g_{n'}}
\ee
\be\label{fbk58}
\Big|{\cal A}_{\Pi^{N}_{\{g_{n}\}},\Pi^{N'}_{\{g_{n'}\}}}\Big|^{2}={\cal X}_{\Pi_{N},\Pi_{N'}} \,\prod_{n}\big(V^{2}_{n}\big)^{g_{n}} \prod_{n'}\big({{V_{n'}'}^{2}}\big)^{g_{n'}}
\ee
it is clear that if one takes $n,n'=1$ and $g_{1}=N,\,g'_{1}=N'$ then $V_{1}=V'_{1}=1$, and even if the polynomial ${\cal P}_{\Pi_{N},\Pi_{N'}}$ is very complicated there are no chaotic oscillations. This is a clear information of how the polynomial ${\cal P}_{\Pi_{N},\Pi_{N'}}$ does not contain most of the information about the microstate structures.
When ${\cal X}_{\Pi_{N},\Pi_{N'}}$ is considered, there is still a complicated structure, but there is a constant universal leading term plus many other suppressed terms. 
The crucial point is that, if there is thermalization the only way to generate the process is using the chaotic factors $\{V_{n}\}, \{V'_{n'}\}$. This is an evidence of how the chaos is a trigger for the thermalization.

To see how the ${\cal X}_{\Pi_{N},\Pi_{N'}}$ contributes, let's start by considering
\be\label{fbk59}
\Big|{\cal P}_{\Pi_{N},\Pi_{N'}}\Big|^{2}={\cal X}_{\Pi_{N},\Pi_{N'}}\left(p'{\cdot}{\cal L}^{(a)}{\cdot}p',\,p{\cdot}{{\cal L}'}^{(a)}{\cdot}p,\,p'{\cdot}{\cal L}^{(a)}{\cdot}{{\cal L}'}^{(b)}{\cdot}p,\, w_{n,m},\,w'_{n',m'},\,\mu_{n,n'} \right)
\ee
where
\be\label{fbk60}
\mu_{n,m}(E_{k})=-{n m {E_{k}\over M_{N}}\over m - n+n{E_{k}\over M_{N}}}\,,\quad w_{n,m}(E_{k})={n\,m\over n+m}  {E_{k}\over M_{N}} \left( {E_{k}\over M_{N}} -1\right)
\ee
\be\label{fbk61}
w'_{n,m}(E_{k})={n\,m\over n+m}  {E_{k}\over M_{N}} \left( {E_{k}\over M_{N}} +1\right)
\ee
when $E_{k}/M_{N}$ is small, which is the region where the emitted radiation is expected to be thermal, one has
\be\label{fbk62}
w'_{n,m}=-w_{n,m}={n\,m\over n+m}  {E_{k}\over M_{N}} +O\left( {E_{k}^{2}\over M_{N}^{2}}\right)
\ee
and the other terms are given by (\ref{fbk39}), therefore if $E_{k}/M_{N}$ is small one has 
\be\label{fbk63}
p'{\cdot}{\cal L}{\cdot}p'=p{\cdot}{\cal L}'{\cdot}p=p'{\cdot}{\cal L}{\cdot}{\cal L}'{\cdot}p\simeq 2=-2\alpha' M_{T}^{2} 
\ee
and the leading term of the polynomial ${\cal X}$, included the microstates normalization is
\be\label{fbk64}
{\cal N}_{\{g_{n}\}} {\cal N}'_{\{g'_{n'}\}} {\cal X}_{\Pi_{N},\Pi_{N'}}\simeq { (-2\alpha' M_{T}^{2}) ^{J+J'}\over \prod_{n} n^{g_{n}} g_{n}!\,\prod_{n'} n'^{g_{n'}} g_{n'}! }+ O\left({E_{k}\over M_{N}} \right)
\ee
so one can absorb the polynomial contribution into a new microstates normalization
\be\label{fbk65}
\widetilde{{\cal N}}_{\{g_{n}\}} \widetilde{{\cal N}}'_{\{g'_{n'}\}}={1\over \prod_{n} (n/2)^{g_{n}} g_{n}!\, \prod_{n'} (n'/2)^{g_{n'}} g_{n'}! }
\ee
Now focusing on the chaotic factors, $V_{n}$ and $V'_{n'}$, when $E_{k}/M_{N}$ is small one can write
\be\label{fbk66}
V_{n}={\Gamma\left(n-n{E_{k}\over M_{N}} \right)\over \Gamma(n)\Gamma\left(1-n{E_{k}\over M_{N}} \right)}
\ee 
Using the Legendre duplication formula
\be\label{fbk67}
\Gamma(n x)={\prod_{k=0}^{n-1}\Gamma\left(x + {k\over n}\right)\over (2\pi)^{n-1\over 2} n^{1\over 2} n^{-n x}  }
\ee
one has
\be\label{fbk68}
\Gamma\left(n\left(1-{E_{k}\over M_{N}}\right) \right)={\prod_{k=0}^{n-1}\Gamma\left(1-{E_{k}\over M_{N}} + {k\over n}\right)\over (2\pi)^{n-1\over 2} n^{{1\over 2} -n} } \,e^{-n{E_{k}\over M_{N}}\log n}
\ee 
since $E_{k}/M_{N}$ is very small one can approximate the function as
\be\label{fbk69}
\Gamma\left(n\left(1-{E_{k}\over M_{N}}\right) \right)=\Gamma(n)\,e^{-n{E_{k}\over M_{N}}\log n}
\ee
and the final result is given by
\be\label{exptherm}
V_{n}\simeq  {\Gamma\left(n-n{E_{k}\over M_{N}} \right)\over \Gamma(n)\Gamma\left(1-n{E_{k}\over M_{N}} \right)}\simeq  {e^{-n{E_{k}\over M_{N}}\log n}\over \Gamma\left(1-n{E_{k}\over M_{N}} \right) }
\ee

In a similar fashion the other term can be written as
\be\label{exptherm2}
V'_{n}=(-)^{n}{\Gamma\left(n+n{E_{k}\over M_{N}} \right)\over \Gamma(n) \Gamma\left(1+n {E_{k}\over M_{N}}  \right) }\simeq (-)^{n}\,  {e^{n{E_{k}\over M_{N}}\log n}\over \Gamma\left(1+n{E_{k}\over M_{N}} \right) }
\ee 
this is how the chaotic factors play their role in the thermalization of the decay rate. 

\subsection{Thermal nature of the decay amplitude}
Let's consider the general form of the decay amplitude
\be\label{fbk70}
{\cal A}_{\Pi^{N}_{\{g_{n}\}},\Pi^{N'}_{\{g_{n'}\}}}= {\cal P}_{\Pi_{N},\Pi_{N'}}\, \prod_{n}\big(V_{n}\big)^{g_{n}} \prod_{n'}\big({V_{n'}'}\big)^{g_{n'}}
\ee
in the limit $E_{k}/M_{N}$ small, it was shown that the decay rate, which is the physical observable, has a simple compact dependence on the polynomial ${\cal X}_{\Pi_{N},\Pi_{N'}}$ that can be absorbed in the microstates normalization. Such simplification can be translated in the structure of the decay amplitude when only circular polarizations are considered, for example taking
\be\label{fbk71}
\lambda_{+}=\lambda'_{+}=(0,1,i,\vec{0})
\ee
one finds 
\be\label{fbk72}
\zeta_{n}{\cdot}p'=\zeta'_{n'}{\cdot}p\simeq-{\omega}\,, \quad \zeta_{n}{\cdot}\zeta_{n}=\zeta'_{n'}{\cdot}\zeta'_{n'}=\zeta_{n}{\cdot}\zeta'_{n'}=0
\ee 
and the decay amplitude simplifies to
\be\label{fbk73}
{\cal A}_{\Pi^{N}_{\{g_{n}\}},\Pi^{N'}_{\{g_{n'}\}}}={\cal N}_{\{g_{n}\}} {\cal N}'_{\{g'_{n'}\}}\left(-{\omega}\right)^{J+J'}\, \prod_{n}\big(V_{n}\big)^{g_{n}} \prod_{n'}\big({V_{n'}'}\big)^{g_{n'}}
\ee
Now using (\ref{exptherm}) and (\ref{exptherm2}) one finds
\be\label{fbk74}
{\cal A}_{\Pi^{N}_{\{g_{n}\}},\Pi^{N'}_{\{g_{n'}\}}}={\cal N}_{\{g_{n}\}} {\cal N}'_{\{g'_{n'}\}}\left(-{\omega}\right)^{J+J'}\,  e^{-{\cal C}_{N}(\{g_{n}\},\{g'_{n'}\}){E_{k}\over 2 T_{H}}-\mu_{N}\left( \{g_{n}\},\{g'_{n'}\};E_{k}/T_{H}\right)}
\ee
and squaring the decay amplitude one recovers the formula of the decay rate (\ref{thermA}). The kinematical simplifications of the decay amplitude are just the reflection of the subleading contributions of bilinear polarization terms of the decay rate, in the limit $E_{k}/M_{N}$ small. The possibility of computing the most general decay amplitude, with the explicit dependence on the microstates structure leads to the identification of the decay amplitude with the Boltzmann factor, which is the thermal weight of the interaction between microstates.

\end{appendix}

\bibliographystyle{utphys}

\end{document}